\documentclass{article}

\usepackage{arxiv}

\usepackage[utf8]{inputenc} 
\usepackage[T1]{fontenc}    
\usepackage{hyperref}       
\usepackage{url}            
\usepackage{booktabs}       
\usepackage{amsfonts}       
\usepackage{nicefrac}       
\usepackage{microtype}      

\usepackage{graphicx}
\usepackage{natbib}
\usepackage{doi}

\usepackage{bm}
\usepackage{caption}

\usepackage{epstopdf, epsfig}
\usepackage{amsthm}

\usepackage{amsmath}
\usepackage{accents} 
\usepackage{amssymb}
\usepackage{amsxtra}
\usepackage{enumitem}
\usepackage{bbold}

\usepackage{subcaption}
\usepackage{upgreek}
\usepackage{url}
\usepackage{cleveref}       

\newcommand{\dd}{\,\mathrm{d}}

\newcommand{\norm}[1]{\left\lVert #1 \right\rVert}

\DeclareMathOperator\supp{supp}

\newtheorem{definition}{Definition}
\newtheorem{theorem}{Theorem}
\newtheorem{lemma}{Lemma}

\newtheorem{postulate}{Postulate}

\crefname{section}{§}{§§}
\Crefname{section}{§}{§§}

\title{On a mathematical definition of laminar and turbulent fluid flow}

\newif\ifuniqueAffiliation
\uniqueAffiliationtrue

\ifuniqueAffiliation 
\author{ \href{https://orcid.org/0000-0002-8065-1449}{\includegraphics[scale=0.06]{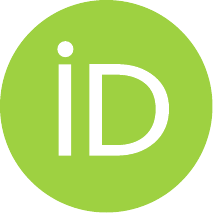}\hspace{1mm}F. Javier Garc\'ia Garc\'ia} \\
	Thermal Systems \& Heat Transfer  \\
	Research Group (SISTER)\\
	University of A Coru\~na\\
	Campus Industrial de Ferrol\\
	C/ Mendiz\'abal, s/n\\
	15403-Ferrol, A Coru\~na,
	Spain\\
	\texttt{f.javier.garcia.garcia@udc.es} \\
	\And
	\href{https://orcid.org/0000-0002-9598-5249}{\includegraphics[scale=0.06]{orcid.pdf}\hspace{1mm}Pablo Fari\~nas Alvari\~no} \\
	Department of Naval \& Industrial Engineering\\
	Polytechnic School of Engineering \\
	University of A Coru\~na\\
	Campus Industrial de Ferrol\\
	C/ Mendiz\'abal, s/n\\
	15403-Ferrol, A Coru\~na,
	Spain\\
	\texttt{pablo.farinas@udc.es} \\
}
\else
\usepackage{authblk}

\newbox{\orcid}\sbox{\orcid}{\includegraphics[scale=0.06]{orcid.pdf}} 
\author[1]{%
	\href{https://orcid.org/0000-0000-0000-0000}{\usebox{\orcid}\hspace{1mm}David S.~Hippocampus\thanks{\texttt{hippo@cs.cranberry-lemon.edu}}}%
}
\author[1,2]{%
	\href{https://orcid.org/0000-0000-0000-0000}{\usebox{\orcid}\hspace{1mm}Elias D.~Striatum\thanks{\texttt{stariate@ee.mount-sheikh.edu}}}%
}
\affil[1]{Department of Computer Science, Cranberry-Lemon University, Pittsburgh, PA 15213}
\affil[2]{Department of Electrical Engineering, Mount-Sheikh University, Santa Narimana, Levand}
\fi

\renewcommand{\headeright}{Scientific Article}
\renewcommand{\undertitle}{Scientific Article}
\renewcommand{\shorttitle}{On a mathematical definition of laminar and turbulent flow}

\hypersetup{
pdftitle={On a mathematical definition of laminar and turbulent fluid flow},
pdfsubject={physics.flu-dyn},
pdfauthor={F.J.~Garcia Garcia, P.~Fari\~nas Alvari\~no},
pdfkeywords={Turbulence, Turbulent flow, Navier-Stokes equation},
}

\begin{document}
\maketitle

\begin{abstract}
As stated in the title, the present research proposes a mathematical definition of laminar and turbulent flows, i.e., a definition that may be used to conceive and prove mathematical theorems about such flows.
The definition is based on an experimental truth long known to humans: Whenever one repeats a given flow, the results will not be the same if the flow is turbulent.
Turbulent flows are not strictly repeatable.
From this basic fact follows a more elaborate truth about turbulent flows: The mean flow obtained by averaging the results of a large number of repetitions is not a natural flow,
that is, it is a flow that cannot occur naturally in any experiment.
The proposed definition requires some preliminary mathematical notions, which are also introduced in the text:
Proximity between functions, the ensemble of realisations, the method of averaging the flows, and the distinct properties of realisations (physical flows) and averages (mean flows).
The notion of restricted nonlinearity is introduced and it is demonstrated that laminar flows can only exist in conditions of restricted nonlinearity, 
whereas turbulent flows are a consequence of general nonlinearity.
The particular case of steady-state turbulent flow is studied, and an uncertainty is raised about the equality of ensemble average and time average.
Two solved examples are also offered to illustrate the meaning and methods implied by the definitions: 
The von K\`arm\`an vortex street and a laminar flow with imposed white-noise perturbation.
\end{abstract}


\

\textbf{Laminar and turbulent flows are essentially different, although both are solutions of the same Navier-Stokes equations (NSE).
While most of their differences are not readily amenable to mathematical analysis, there is one that may be used to distinguish between them: 
Laminar flows are repeatable. 
This essential difference warrants a mathematical definition which, to the best of our knowledge, is the first ever published.}

\section{Introduction}\label{sct:intro}

The study of turbulence from a scientific-mathematical point of view has been seamlessly developed over at least one and a half centuries.
There are more than forty books whose sole subject is the study of turbulence as a natural physical phenomenon.
In general, they are profound texts that demand a thoughtful and focused effort from readers, and some of them would plausibly be considered classic books.
Those books explain at length the main characteristics of turbulent flows and, invariably, they all emphasise how difficult it is to devise a definition of turbulence.
Some of them even offer a textual (most often quite eloquent) definition of turbulence and, particularly, \citep[App. A]{Tsi09} compiles an interesting catalogue of descriptive definitions proposed by various authors.
However, none of them venture a mathematical definition of a turbulent flow.
Arguably, such a definition could be deduced from the contents of the books, since they expose the mathematical apparatus needed to understand turbulence, if ever someone can.
But a statement of the sort ``... \textit{a turbulent flow is defined as} ...'', followed by a mathematical proposition, seems to be lacking in the extensive literature on turbulence.

There exists a general intuitive notion of what a turbulent flow is, gathered along many decades, mostly derived from the chaotic motion that is observed in any realisation.
Most specialists in fluid mechanics, if requested to propose a definition of turbulent flow, would mention the fluctuating components, the vortical structure, 
the many scales of motion from inertial range to Kolmogorov microscale, the energy cascade and vortex stretching, the viscous dissipation at the smallest scale, 
and many other features that may be directly perceived during any experiment.
But if they were requested to propose a mathematical statement, rather than descriptive words, possibly most specialists would find it hard to select a mathematical proposition that would define a turbulent flow. 

It turns out that an unambiguous mathematical definition of turbulent flow is possible, and it is not very far from the conventional notion that,
more or less, most specialists have of it.
The proposed definition does not rely on the specific properties that may be observed in each realisation, but rather on the aggregate behaviour of nearly-identical instances of the same flow.
Despite the complexity of the mathematical apparatus required to enunciate the definition, the basic idea is quite simple and intuitive, and arguably most readers would be amenable to accepting it after a brief reflection.
A somewhat similar notion can be found in \citep[Sect. 5.1]{MY71}, where it is asserted that ''...\textit{we shall define as turbulent only those flows for which there exists a statistical ensemble of similar flows, characterized by some probability distribution (with continuous density) for the values of all possible fluid dynamic variables}...``.
However, we do not need to assume a probability distribution for the fields that occur in the realisations; 
it is an unnecessary hypothesis for our proposed definition.

Our definition of turbulent flow is a natural evolution of the ideas we have been developing along our previous publications (e.g. \citep{GF22}).
The proposed definition makes use of single-point statistics, being unnecessary the recourse to more complicated statistics,
and is valid for general flows, not being limited to steady-state flows.
The present version applies to incompressible flows, although it can be readily extended to compressible flows.
Instead of focusing on the extraordinary dynamical properties of an actual turbulent flow, our approach is based on a less remarkable feature:
Whenever one tries to repeat a turbulent flow, the results obtained are never quite the same.
We believe this exclusive attribute of turbulent flows (and many other chaotic dynamical systems) is the most suitable to be used in a mathematical definition,
because to be or not to be the same is easily translated into mathematical language.
To evaluate such a property, a notion of proximity among fields must be established.
Two flows are considered the same if their respective fields are sufficiently proximate along their entire evolution. 
The mathematical definition of turbulent flow arises naturally when an ensemble of should-be-equal flows exists, in which the proximity among fields can be calculated. 
 
This article offers a perspective of turbulent flows not based upon their intrinsic properties, but rather on the collective behaviour of many realisations that begin in (almost) identical conditions.
The key point resides on the repeatability of the resulting flows, being laminar those that are repeatable.
It is an experimental truth that turbulent flows have a distinct average behaviour, neatly different from that of laminar flows. 
It should be stressed that other definitions are also possible, perhaps even more refined than ours.
Hopefully, after reading our approach to this problem, some researchers will eventually come up with other mathematical ways of defining turbulent flows.

This research is complemented with four appendices.
A short non-mathematical summary of the proposed definitions is offered in \cref{sct:simpleIntro}.
Also, two solved examples in \cref{sct:KarmanVortex} \& \cref{sct:whiteNoise} will help understand the meaning and methods of the proposed definitions.
In particular, the example of \cref{sct:whiteNoise} overcomes the apparent contradiction of a laminar flow with imposed white-noise fluctuations.
Finally, \cref{sct:haps} studies how the particular features of a turbulent flow are transmitted to the mean flow obtained after many realisations.

\section{Nomenclature and conventions}\label{sct:nomencl}
This research intends to be part of physics, not mathematics, although its content is highly mathematical. 
Some results can be readily translated to higher dimensions and, possibly, can also be satisfied with less demanding conditions than those herein requested.
However, it is our intention to keep the argumentation as close as possible to the physical reality of fluid flows, rather than pushing the mathematical subtleties to the limit.

Let $\Updelta \subset \mathbb{R}^3$ be the region of the three-dimensional Euclidean space occupied by a flow.
It is assumed that $\Updelta$ is bounded, that is, $\upmu_3(\Updelta) < +\infty$ for any measure
\footnote{In general, we shall take the canonical measure in $\mathbb{R}^n$, $\upmu_n:\mathcal{R}_n \rightarrow \{0 \} \cup  \mathbb{R}^+ $, in which $\mathcal{R}_n$ is a $\sigma$-algebra on $\mathbb{R}^n$, 
and for $n=3$ $\upmu_3(\Updelta)$ is equal to the volume of the domain $\Updelta$.
Also, the canonical measure $\upmu_4:\mathcal{R}_4 \rightarrow  \{0 \} \cup \mathbb{R}^+$ will be considered for spacetime domains.
Throughout the text, the symbol $\upmu_4$ is also written as $\upmu$, $\upmu(\Upomega)\equiv\upmu_4(\Upomega)$, to ease the notation. 
The notion of measure of a domain will be extensively used in the present research.}
$\upmu_3$ in $\mathbb{R}^3$.
Furthermore, $\Updelta$ is assumed a connected open domain with a piecewise-smooth boundary $\partial \Updelta$, being its closure $\overline{\Updelta}=\Updelta \cup \partial \Updelta$.
For instance, $\Updelta$ could be the test section of an experimental rig or the computational domain of a CFD simulation (excluding the boundaries).
Let $I =[0,T_r]\subset \mathbb{R}$ be the bounded time interval during which the flow is being studied; without loss of generality it is assumed to begin at $t=0$.
Let $\Upomega = I \times \Updelta$ be the spacetime bounded domain wherein the flow develops, with spatial boundary $\partial \Updelta$, with $\partial \Upomega = I \times \partial \Updelta$ and with bounded closure $\overline{\Upomega}=\Upomega \cup \partial \Upomega$.
We purposely avoid the consideration of unbounded domain, since it brings an additional set of unnecessary problems.
Let $f:\Upomega \rightarrow \mathbb{R}$, $f(t,\textbf{x})$, be a scalar function of interest in the flow (pressure, temperature, species concentration, etc.),
or the component of a vector or tensor field in the flow (velocity, pressure gradient, vorticity, heat flux, etc.).

The spacetime variables are written interchangeably as $\undertilde{\textbf{\textit{x}}}=(t,\textbf{x})=(x_0,x_1,x_2,x_3)=(x_0,x_i)=(x_{\alpha})$, 
and also $\dd\Upomega = \dd t \dd^3 \textbf{x}=\dd^4 \undertilde{\textbf{\textit{x}}} $.
The derivatives are written interchangeably as:
\begin{equation}
 \label{eq:derivatives}
 \frac{\partial f}{\partial t} =\partial_0 f = f_{,0} \ , \quad
 \frac{\partial f}{\partial x_i} = \partial_i f=f_{,i} \quad (i=1,2,3) \ , \quad
 \frac{\partial f}{ \partial x_{\alpha}}= \partial_{\alpha} f = f_{,\alpha} \quad (\alpha=0,1,2,3)
\end{equation}
and 
\begin{equation}
 \label{eq:2derivatives}
  \frac{\partial^2 f}{\partial x_{i} \partial x_{j}} = \partial_i \partial_j f = \partial^2_{ij}f=f_{,ij} \quad (i,j=1,2,3) \ , \quad
 \frac{\partial^2 f}{\partial x_{\alpha} \partial x_{\beta}} = \partial_{\alpha} \partial_{\beta} f = \partial^2_{\alpha \beta}f= f_{,\alpha\beta} \quad (\alpha,\beta=0,1,2,3)
\end{equation}

Einstein summation convention over repeated indices is assumed throughout this work.
The coordinates on the boundary $\partial \Updelta$ are denoted by $\textbf{s}\mathop{=}(s_1,s_2,s_3) \mathop{\in} \partial \Updelta$, 
being $\dd \Upsigma$ an infinitesimal surface element within $\partial \Updelta$ and $(t,\textbf{s})=\undertilde{\textbf{\textit{s}}} \mathop{\in} \partial \Upomega \mathop{=} I \mathop{\times} \partial \Updelta$.

Functions considered in this research are real-valued functions which belong to a Sobolev functional space $W^{2,2}(\Upomega) \mathop{\equiv} H^2(\Upomega)$, endowed with the inner product
\begin{equation}
 \label{eq:innerProduct}
 \langle f,g \rangle := \int \limits_{\Upomega} \left[ f(t,\textbf{x}) g(t,\textbf{x}) +f_{,\alpha}(t,\textbf{x}) g_{,\alpha}(t,\textbf{x}) +f_{,\alpha\beta}(t,\textbf{x}) g_{,\alpha\beta}(t,\textbf{x}) \right] \dd t \dd^3 \textbf{x} \ < + \infty \ , \quad \forall \ f,g \in H^{2}(\Upomega)
\end{equation}
which grants a structure of Hilbert space to $H^2(\Upomega)$.
This inner product induces a norm in $H^{2}(\Upomega)$ given by:
\begin{equation}
 \label{eq:norm}
 \norm{f} := \langle f,f \rangle^{\tfrac12} = \left \{ \int \limits_{\Upomega} \left [ |f(t,\textbf{x})|^2 +  |f_{,\alpha}(t,\textbf{x})f_{,\alpha}(t,\textbf{x})|  + |f_{,\alpha \beta}(t,\textbf{x})f_{,\alpha \beta}(t,\textbf{x})| \right] \dd \Upomega \right \}^{\tfrac12} < + \infty \ , \quad \forall \ f \in H^{2}(\Upomega)
\end{equation}
which, in turn, permits to define a distance between functions, $\norm{f-g}$.

The support of $f$, written $\supp(f)$, is the closure of the set of points $(t,\textbf{x}) \in \Upomega$ such that  $f(t,\textbf{x})\neq 0$:
\[ \supp(f) := \overline{ \{ (t,\textbf{x})\in \Upomega \ | \ f(t,\textbf{x}) \neq 0 \} }   \]
The indicator function (or characteristic function) of a set $\upomega \subset \Upomega$ is defined by:
\begin{align}
\label{eq:indicatorFunct}
\mathbb{1}_{\upomega}(t,\textbf{x})\mathop{:=}\begin{cases}
1 \ \ &if \ \ (t,\textbf{x})\in \upomega \\
0 \ \ &if \ \ (t,\textbf{x})\notin \upomega \\
\end{cases}
\end{align}
The indicator function $\mathbb{1}_{\upomega}(t,\textbf{x})$ takes identically the value $1$ at any point of its support.

\section{Physical functions of range \(\epsilon\)}
\label{sct:physFunc}
The definition of turbulent flow that will be introduced later is related with repeating the same experiment many times and performing statistical calculations with the obtained results.
The experiments should be as identical as physically possible (but not mathematically identical) and this requires establishing criteria to evaluate the proximity of functions, 
be them solutions of governing equations, initial or boundary conditions, source terms or of any other type.
Thus, first we have to introduce the notion of approximate functions and establish criteria to assess when two given functions can be considered almost equal.
The idea of approximate functions is necessary to correctly address the extreme dependency on initial and boundary conditions (I\&BC), typical of non-linear systems.

\begin{definition}
\label{def:almostEq}
Let $f,g\in H^{2}(\Upomega)$ be real-valued functions and $\epsilon \in \mathbb{R}, \ \epsilon >0$, an arbitrarily small number.
The functions $f,g$ are \textbf{quasi-equal of range} $\bm{\epsilon}$, denoted $f\cong_{\epsilon} g$, iff the following conditions are satisfied:
\begin{enumerate}[label=(\arabic*)]
\item There exists an open set $\Upomega' \subset \Upomega$, with $\upmu(\Upomega') \neq 0$, such that $f(t,\textbf{x}) \neq g(t,\textbf{x}) \  \forall \ (t,\textbf{x})present \in \Upomega'$. 
\item
\begin{align}
 \label{eq:almostEqual0}
&\norm{f-g} = \langle f-g,f-g \rangle^{\tfrac12} =  \\
&\left \{ \int \limits_{\Upomega} \left[ |f(t,\textbf{x})-g(t,\textbf{x})|^2 + |f_{,\alpha}(t,\textbf{x})-g_{,\alpha}(t,\textbf{x})|^2 +  |f_{,\alpha\beta}(t,\textbf{x})-g_{,\alpha\beta}(t,\textbf{x})|^2 \right] \dd \Upomega \right \}^{\tfrac12} < \epsilon  \nonumber
\end{align}
\end{enumerate}
\end{definition}
Such functions $f,g$ are not only themselves very approximate during the temporal evolution of the system, but also all their first and second derivatives.
Condition (1) above prevents $f$ and $g$ from being mathematically identical.
This will be the general notion of approximate functions considered in this research, mainly because it considers the cumulative difference between them, 
i.e., it measures the level and extension of the difference between both functions.

When functions do not depend upon time, it is convenient a notion of proximity that is only concerned with their spatial behaviour.
This approach is useful to define approximate initial conditions for differential equations, as it will be seen below.
\begin{definition}
\label{def:spatEqual}
Let $f,g \in H^2(\Updelta)$ be real-valued functions and $\epsilon \in \mathbb{R}, \ \epsilon >0$, an arbitrarily small number.
The functions $f,g$ are \textbf{spatially quasi-equal of range} $\bm{\epsilon}$, denoted $f \simeq_{\epsilon} g$, iff the following conditions are satisfied:
\begin{enumerate}[label=(\arabic*)]
\item There exists an open set $\Updelta' \subset \Updelta$, with $\upmu_3(\Updelta') \neq 0$, such that $f(\textbf{x}) \neq g(\textbf{x}) \  \forall \ \textbf{x} \in \Updelta'$.
\item
\begin{equation}
 \label{eq:almostSpaEqual0}
\left \{ \int \limits_{\Updelta} \left[ |f(\textbf{x})-g(\textbf{x})|^2 + |f_{,i}(\textbf{x})-g_{,i}(\textbf{x})|^2 + |f_{,ij}(\textbf{x})-g_{,ij}(\textbf{x})|^2 \right] \dd^3 \textbf{x}  \right \}^{\tfrac12} < \epsilon 
\end{equation}
\end{enumerate}
\end{definition}

Now, the notion of proximity will be defined for the boundary of the system.
It is recalled that $\partial \Updelta$ is the piecewise-smooth boundary of the spatial domain $\Updelta$ and $\partial \Upomega = [0,T_r] \times \partial \Updelta$.
It will be seen below that the following definition is useful to set approximate boundary conditions for partial differential equations:
\begin{definition}
\label{def:boundaryEqual}
Let $f,g \in H^2(\partial\Upomega)$ be real-valued functions and $\epsilon \in \mathbb{R}, \ \epsilon >0$, an arbitrarily small number.
The functions $f,g$ are \textbf{boundary quasi-equal of range} $\bm{\epsilon}$, denoted $f \asymp_{\epsilon} g$, iff the following conditions are satisfied:
\begin{enumerate}[label=(\arabic*)]
\item There exists an open set $\partial \Upomega' \subset \partial \Upomega$, with $\upmu_3(\partial \Upomega') \neq 0$, such that $f(t,\textbf{s}) \neq g(t,\textbf{s}) \  \forall \  (t,\textbf{s}) \in \partial \Upomega'$.
\item
\begin{equation}
 \label{eq:almostBoundEqual0}
\left \{ \int \limits_{\partial \Upomega} \left[ |f(t,\textbf{s})-g(t,\textbf{s})|^2 + |f_{,\alpha}(t,\textbf{s})-g_{,\alpha}(t,\textbf{s})|^2 + |f_{,\alpha\beta}(t,\textbf{s})-g_{,\alpha\beta}(t,\textbf{s})|^2 \right] \dd t \dd \Upsigma  \right \}^{\tfrac12} < \epsilon 
\end{equation}
\end{enumerate}
\end{definition}

Usually, we shall say that functions are $\epsilon$-equal or $\epsilon$-identical to denote any of the three quasi-equalities of range $\epsilon$ defined above.
The reader may wonder why it has been chosen an approximation based on the integrals above instead of other possibilities, such as the following pointwise inequality:
\begin{equation}
\label{eq:poinwise}
 |{f}(t,\textbf{x})-{g}(t,\textbf{x}) | < {\epsilon}, \ \forall \ (t,\textbf{x}) \in \Upomega 
 \end{equation}
and analogous inequalities for the derivatives.
Three main reasons support our choice: (i) the integral is the canonical norm of a Sobolev space, (ii) the integral measures the cumulative difference between two $\epsilon$-equal functions,
not just a set of local differences, and (iii) for given $\epsilon$, if $\upmu(\Upomega)>1/21$ then the integral (which has 21 terms) is in general a finer approximation than pointwise inequalities when derivatives are included,
except perhaps in a small domain $\Upomega_0 \subset \Upomega$ (recall the domain $\Upomega$ is bounded).

The notion of $\epsilon$-equal functions brings forth the following new concept:
\begin{definition}
 \label{def:physFunct}
 Let $f\in H^{2}(\Upomega)$ be a real-valued function and $\epsilon \in \mathbb{R}, \ \epsilon >0$, an arbitrarily small number.
The function $f$ induces a \textbf{physical function of range} $\bm{\epsilon}$, defined as the set of functions $f_{[\epsilon]}=\{ g\in H^{2}(\Upomega) \ |\ g\cong_{\epsilon} f \}$.
\end{definition}
Any two functions $g_1,g_2 \in f_{[\epsilon]}$ are quasi-equal of range $2\epsilon$ between them.
The expression $f_{[\epsilon]} (t,\textbf{x})$ will denote any function $g \in f_{[\epsilon]}$, including $f(t,\textbf{x})$ itself.
The relationship "$g_1 \sim g_2$ \textit{ iff} $g_1\mathop{\cong_{\epsilon}}f$ \textit{and} $g_2\mathop{\cong_{\epsilon}}f$" is of equivalence and creates the equivalence class  $f_{[\epsilon]}(t,\textbf{x})$; 
any function $ g \in f_{[\epsilon]}$ represents equally well the equivalence class $f_{[\epsilon]}(t,\textbf{x})$.
We shall say that a class $f_{[\epsilon]}(t,\textbf{x})$ is the solution of a given mathematical expression (algebraic or differential equation),
if there exists at least one function $g \in f_{[\epsilon]}$ in the class that satisfies the mathematical expression. 
It follows that other functions of the class $f_{[\epsilon]}(t,\textbf{x})$ will only approximately satisfy the given mathematical expression,
although the approximation could be as fine as desired.
Obviously, an equivalence class become meaningless if $\epsilon$ is chosen too large.

The above definition purports to address the following situation.
Imagine that we have a measuring instrument whose intrinsic uncertainty is greater than $\epsilon$.
Two measurements differing in less than $\epsilon$ will be recorded by the instrument as the same measurement, they will be indistinguishable.
Therefore, if we say "\textit{we have measured} $f_{[\epsilon]} (t,\textbf{x})$ \textit{in the experiment}", 
what is really meant is that we would have obtained any function $g \in f_{[\epsilon]}$, including $f(t,\textbf{x})$ itself.
Thus, the physical function of range $\epsilon$ appears to be a single function to the experimental researcher, always provided that $\epsilon$ be smaller than the measurement uncertainty of the corresponding instrument.

The physical function of range $\epsilon$ has two very interesting general properties:
\begin{enumerate}
 \item The set $f_{[\epsilon]}$ has the cardinality of the continuum. 
 That is to say, there are non-countable infinite functions within $f_{[\epsilon]}$ and the distance between any two functions $g_1,g_2 \in f_{[\epsilon]}$ is seamlessly covered with functions belonging to  $f_{[\epsilon]}$. 
\item There is an infinite subset of polynomial functions within  $f_{[\epsilon]}$ that is dense in  $f_{[\epsilon]}$, i.e., any function  $g \in f_{[\epsilon]}$ is arbitrarily approximate to a polynomial function.
This property allows to use simple polynomial functions as suitable approximations to any function in $f_{[\epsilon]}$. 
\end{enumerate}

The notion of physical functions of range $\epsilon$, when applied to the steady-state case, i.e., to time-independent functions, 
brings forth a set of $\epsilon$-identical stationary functions $f:\Updelta \rightarrow \mathbb{R}$,
which can be employed as initial conditions for differential equations.
The following definition clarifies this notion.
\begin{definition}
\label{def:physInitCond}
  Let $f\in H^{2}(\Updelta)$ be a real-valued function and $\epsilon \in \mathbb{R}, \ \epsilon >0$, an arbitrarily small number.
The function $f$ induces a \textbf{physical initial condition of range} $\bm{\epsilon}$, defined as the set of functions $f_{\langle \epsilon \rangle}=\{ g\in H^{2}(\Updelta) \ |\ g \simeq_{\epsilon} f \}$.
\end{definition}
For $\epsilon$ sufficiently small, any function $g \in f_{\langle \epsilon \rangle}$ would serve as a suitable initial condition for the physical problem,
since no measuring instrument would record the difference.
As stated above, the expression $f_{\langle \epsilon \rangle} (\textbf{x})$ will denote any initial condition $g\in f_{\langle \epsilon \rangle}$, 
including $f(\textbf{x})$ itself, and constitutes an equivalence class.

Likewise, the notion of physical function is also applied to the boundary of a dynamical system, bringing forth the following definition for a set of equivalent boundary conditions:
\begin{definition}
\label{def:physBoundCond}
  Let $f\in H^{2}(\partial\Upomega)$ be a real-valued function and $\epsilon \in \mathbb{R}, \ \epsilon >0$, an arbitrarily small number.
The function $f$ induces a \textbf{physical boundary condition of range} $\bm{\epsilon}$, defined as the set of functions $f_{ \ulcorner \epsilon \lrcorner}=\{ g\in H^{2}(\partial\Upomega) \ |\ g \asymp_{\epsilon} f \}$.
\end{definition}
The symbol $f_{ \ulcorner \epsilon \lrcorner}$ does not only refer to the canonical boundary conditions of Dirichlet (function at the boundary), 
Neumann (normal derivative at the boundary) or Robin (linear combination of both), 
but also to general conditions affecting any derivative at the boundary or any combination of the above in different patches of the boundary $\partial \Updelta$.
The key point is that any member of the equivalence class $f_{ \ulcorner \epsilon \lrcorner } (t,\textbf{s})$ would represent the same boundary conditions within $\epsilon$.
Note the functions of $f_{ \ulcorner \epsilon \lrcorner } $ may have different physical dimensions, i.e., being measured with different units, in different boundary patches,
since one patch could comprise the function itself while another may involve its derivative.

\section{The ensemble of realisations}%
\label{sct:ensembAve}

The notions of realisation, ensemble and ensemble-averaged field have already been addressed in \citep[Sect. 3.2]{Gar17}, but a renewed account will be offered herein, with some new insight.
Let us consider an experimental rig for a given type of flow, in which a source of pressure gradient is connected to it and the gravity acts upon the system. 
For the moment, these are the only external forces applied and we ignore other dynamical effects.
The open connected set $\Updelta \subset \mathbb{R}^3$, $ 0 < \upmu_3(\Updelta) < +\infty$, with piecewise-smooth boundary $\partial \Updelta$ considered so far, would be the bounded spatial domain occupied by the rig's test section.
Assume all experiments are carried out exactly during the finite time interval $I=[0,T_r]$, $T_r < +\infty$, with no loss of generality for taking the initial instant as zero.
Let $\Upomega = [0,T_r] \times \Updelta \subset \mathbb{R}^4$ be the connected spacetime domain in which every experiment develops, 
$\partial \Upomega = [0,T_r] \times \partial \Updelta \subset \mathbb{R}^4$ and bounded closure $\overline{\Upomega} = \Upomega \cup \partial \Upomega$.
Consider a physical flow field $\psi \mathop{\in} H^2(\Upomega)$ (velocity, pressure, temperature, vorticity...) defined within the domain $\Upomega$.
Suppose we have an all-encompassing set of instruments that can measure $\psi(t,\textbf{x}), \ (t,\textbf{x}) \in \Upomega$, with great accuracy everywhere, being $\epsilon$ the cumulative measurement uncertainty
(or cumulative computer-simulation error).

We plan to execute an arbitrarily large number of as-identical-as-possible experiments.
The initial conditions of $\psi$ at $t=0$ for each experiment would be selected within the equivalence class constituted by the physical initial conditions of range $\epsilon$,
$\psi_{0\langle \epsilon \rangle}(\textbf{x})$ (named after the function $\psi_0(\textbf{x})$, see Def. \ref{def:physInitCond}), and likewise the boundary conditions belong to a physical boundary condition of range
\footnote{It is perfectly possible to generalise the problem by defining a physical initial condition of range $\epsilon_1$, $\psi_{0\langle \epsilon_1 \rangle}(\textbf{x})$,
and a physical boundary condition of range $\epsilon_2$, $\psi_{\partial \ulcorner \epsilon_2 \lrcorner}(t,\textbf{s})$, but such a generalisation would not contribute any substantial insight into the present research.}
$\epsilon$, $\psi_{\partial \ulcorner \epsilon \lrcorner}(t,\textbf{s})$
(named after the function  $\psi_{\partial}(t,\textbf{s})$, see Def. \ref{def:physBoundCond}).
The function $\psi_{\partial}(t,\textbf{s})$ can have dimensions of $\psi$ in some patches of the boundary and also dimensions of $\psi_{,\alpha}$ in other patches of the boundary;
it is only requested that the class $\psi_{\partial \ulcorner \epsilon \lrcorner}$ contains the complete set of boundary conditions demanded by the physical problem.

Let us execute the flow experiment once, with any of the initial conditions in $\psi_{0\langle \epsilon \rangle}(\textbf{x})$ and boundary conditions in $\psi_{\partial \ulcorner \epsilon \lrcorner}(t,\textbf{s})$.
We shall call first \textbf{realisation} to the resulting flow.
The field obtained after this process will be denoted by $\psi^{(1)}(t,\textbf{x})$.
Let us run the experiment a second time, making sure the I\&BC are as identical as physically possible to the previous experiment,
always within $\psi_{0\langle \epsilon \rangle}(\textbf{x})$ and $\psi_{\partial \ulcorner \epsilon \lrcorner}(t,\textbf{s})$.
The result of this second realisation of the flow would be $\psi^{(2)}(t,\textbf{x})$.
Repeat the process indefinitely, always with extreme care of matching the I\&BC as best as possible, always within $\psi_{0\langle \epsilon \rangle}(\textbf{x})$ and $\psi_{\partial \ulcorner \epsilon \lrcorner}(t,\textbf{s})$,
obtaining in each case the field $\psi^{(n)}(t,\textbf{x})$, where the index $n=1,2,...,N$ denotes the realisation.
Note each $\psi^{(n)}(t,\textbf{x})$ is a physical field, that is, a physically realisable field that has, at least, occurred once in history.
The set of realisations $\{ \psi^{(n)}(t,\textbf{x}) \}$, generated from $\psi_{0\langle \epsilon \rangle}(\textbf{x})$ and $\psi_{\partial \ulcorner \epsilon \lrcorner}(t,\textbf{s})$,
is called the \textbf{ensemble} and is denoted by \(\mathfrak{E}_{\epsilon}\) or simply \(\mathfrak{E}\) if the range of proximity $\epsilon>0$ is clear from the context.
Since all realisations start from $\epsilon$-identical initial conditions $\psi_{0\langle \epsilon \rangle}(\textbf{x})$, it is expected that $\psi^{(n)}(0,\textbf{x}) \simeq \psi^{(m)}(0,\textbf{x})$ to a high degree,
for any two realisations $n$ and $m$, and also that $\psi^{(n)}(t,\textbf{x}) \approx \psi^{(m)}(t,\textbf{x})$ to a lesser degree for $t  \approx 0$.
As $t$ increases, the fields  $\psi^{(n)}(t,\textbf{x})$ and $\psi^{(m)}(t,\textbf{x})$ may begin to depart significantly (see Postulates \ref{pos:laminar} \& \ref{pos:turbulent} in \cref{sct:defLamTurb}).

A key aspect in the generation of the ensemble is the random choice of the I\&BC for each realisation; the selection process is inherently stochastic and the composition of the ensemble is randomly built.
In this regard, the notion of ensemble proposed herein is somewhat different from the ''\textit{statistical ensemble of all similar flows}`` requested in \citep[Sect. 3.2]{MY71}.
We do not demand that all flow realisations evolve similarly; we simply request that they all begin similarly and are subject to alike boundary conditions and external forces.
Afterwards, each realisation would behave as demanded by its own dynamics.
The bounded and infinite variability of $\epsilon$-equal I\&BC, randomly chosen, is an essential aspect of the present theory, so that a well-populated ensemble of flows may be generated.
However, $\epsilon$ should be sufficiently small for any suitable ensemble, otherwise the realisations could not be classified as representatives of the same type of flow.
Likewise, the time $T_r$ should be sufficiently long for the flow to evolve significantly.
Some results of the present research would not make sense for large $\epsilon$ or short $T_r$.

The procedure described above yields an ensemble of realisations for any $N \mathop{\in} \mathbb{N}$ (let us assume for the moment that $\epsilon \mathop{\in} \mathbb{R}^+$ remains unchanged).
In principle, different values of $N$ would generate different ensembles, which may lead to varying ensemble properties that would render ambiguous the very process of creating ensembles.
However, as $N \rightarrow \infty$, the particularities of each ensemble become negligibly small and, in the limit, all ensembles tend asymptotically to the same canonical ensemble that summarises the actual mean properties of the flow, for given $\epsilon>0$.
Such a canonical ensemble is not ambiguous, regardless of the particular sequence of realisations chosen to create it.
As long as $N$ is large enough (and $\epsilon$ small enough), the resulting ensemble may be considered a suitable approximation to the canonical ensemble.  
The general procedure to create an ensemble $\mathfrak{E}$ requires the previous setting of $\psi_{0\langle \epsilon \rangle}(\textbf{x})$ and $\psi_{\partial \ulcorner \epsilon \lrcorner}(t,\textbf{s})$.
It may well occur that boundary conditions be not subjected to variation, e.g., in case of internal flow in contact with static rigid walls.
In such cases, see \citep[footnote in page 147]{Gar17}, the no-penetration condition ($\textbf{n} \cdot \textbf{u}(t,\textbf{s}) =0$, $\textbf{n}$ the unit vector normal to the wall) 
and the no-slip condition ($\textbf{n} \wedge \textbf{u} (t,\textbf{s})=0$) lead invariably to fixed boundary conditions ($\textbf{u}(t,\textbf{s})=0$),
which preclude the use of physical boundary conditions of range $\epsilon$.
Whenever the physical boundary conditions of range $\epsilon$ are not applicable, the ensemble should be generated using only physical initial conditions of range $\epsilon$.
The resulting ensemble, for $N$ sufficiently large and $\epsilon$ sufficiently small, would still represent faithfully the mean properties of the flow.

Additionally, flows may be subject to prescribed external body forces, $\textbf{F}(t,\textbf{x})$, which typically appear at the right-hand-side of the Navier-Stokes equation (external sources of motion).
In such cases, selecting randomly the members $\textbf{F}_{[\epsilon]}(t,\textbf{x})$ of a physical body force of range $\epsilon$ in each realisation (see Def. \ref{def:physFunct}), 
would contribute to generate an ensemble $\mathfrak{E}$ of richer variability (gravity is excluded from this procedure, for it may be considered strictly constant).
Those body forces would not fit into the category of initial nor boundary conditions but, if they exist, they must be accounted for when executing each realisation.
Henceforth, in order to simplify this research, no body force other than gravity will be considered and no mention will be made of any equivalence class $\textbf{F}_{[\epsilon]}(t,\textbf{x})$ in coming developments.

Let $\mathfrak{F}_{\Upomega}$ denote the space of physical fields associated to a flow in $\Upomega$, albeit this space will be rigorously defined in \cref{sct:defLamTurb}.
Note that $\mathfrak{E} \subset \mathfrak{F}_{\Upomega}$.
Define now the \textbf{ensemble average} of the physical field $\psi(t,\textbf{x}) \in \mathfrak{F}_{\Upomega}$, for the series of experiments that generate the ensemble $\mathfrak{E}$ described above,
as the mathematical field obtained from one of the following three procedures:
\begin{enumerate}
 \item \textbf{EQUIPROBABLE REALISATIONS}. If all realisations of the ensemble \(\mathfrak{E}\) are equally probable,
 i.e., if selecting one I\&BC from $\{\psi_{0\langle \epsilon \rangle}; \psi_{\partial \ulcorner \epsilon \lrcorner}\}$ is as likely as any other, 
 then the ensemble average of the field $\psi(t,\textbf{x})$ is defined by:
\begin{equation}
 \label{eq:ensembleAve}
 \langle \psi(t,\textbf{x}) \rangle \equiv  \langle \psi \rangle (t,\textbf{x}) \equiv \Psi(t,\textbf{x}) := \lim \limits_{N \rightarrow \infty}
 \frac{1}{N} \sum \limits_{n=1}^{N}  \psi^{(n)}(t,\textbf{x}) \ , \qquad (t,\textbf{x}) \in \Upomega 
\end{equation}
whereby the series is assumed absolutely convergent.
The overwhelming majority of flows satisfy the condition of producing equiprobable realisations; there is no preferred I\&BC within  $\{\psi_{0\langle \epsilon \rangle}; \psi_{\partial \ulcorner \epsilon \lrcorner}\}$.
One cannot guess beforehand which particular field $\psi(t,\textbf{x})$ will arise in a realisation, because one cannot set with absolute mathematical accuracy the I\&BC of each realisation.
 \item \textbf{NOT EQUIPROBABLE REALISATIONS}. Let the probability of producing a type of field $\psi^{(n)}(t,\textbf{x})$ in the $n^{th}$ realisation be $0\leq w_n\leq1$, with a probability distribution $\{ w_n \}$ satisfying
 \begin{equation}
   \lim \limits_{N \rightarrow \infty} \sum \limits_{n=1}^{N} w_n =1 
 \end{equation}
In other words, not all I\&BC within the classes $\{\psi_{0\langle \epsilon \rangle} ; \psi_{\partial \ulcorner \epsilon \lrcorner} \}$ are equally likely to be chosen for a realisation.
In this case, the ensemble average of the field $\psi(t,\textbf{x})$ is defined by:
\begin{equation}
 \label{eq:ensembleAveNEQ}
 \langle \psi(t,\textbf{x}) \rangle \equiv  \langle \psi \rangle (t,\textbf{x}) \equiv \Psi(t,\textbf{x}) := \lim \limits_{N \rightarrow \infty}
  \sum \limits_{n=1}^{N}  w_n \psi^{(n)}(t,\textbf{x}) \ , \qquad (t,\textbf{x}) \in \Upomega 
\end{equation}
whereby the series is assumed absolutely convergent.
The category of equiprobable realisations corresponds to the particular case in which $w_n=1/N, \ \forall \ n \in \mathbb{N}$.
This general case of $w_n\neq 1/N$ is mentioned for the sake of completeness, since we do not know of any practical flow that would fit into such category.
 \item \textbf{CHIRAL FLOWS}. The name of chiral flow is endowed to those flows that have two possible mutually-exclusive configurations, in principle equally probable. 
 Within this category may be included flows such as
 \footnote{Chiral flows show stable laminar configurations for low Reynolds number, which may become turbulent upon increasing $Re$.}:
 The von K\`arm\`an vortex street (in which the first vortex can either be shed at starboard or port side of the cylinder, and subsequent vortices follow the corresponding pattern),
 the Taylor-Couette flow with cell vortices (in which the first vortex could either be clockwise or anticlockwise and the remaining vortices adjust accordingly), 
 the droplet (or very thin jet) falling on top of a horizontal cylinder and dripping from the bottom after sliding over either the left or right semicircular side of the cylinder, etc.
In all those cases, the resulting physical field could either be $\psi^{(n)}_L(t,\textbf{x})$ or $\psi^{(n)}_R(t,\textbf{x})$, ``L'' and ``R'' meaning left and right, respectively.
In principle, $\psi^{(n)}_L(t,\textbf{x})$ and $\psi^{(n)}_R(t,\textbf{x})$ are equally likely, but they should not be averaged together because $(\psi^{(n)}_L +\psi^{(n)}_R)/2$ is nowhere near any one of them.
The procedure for averaging chiral flows is based upon separating the left and right realisations and averaging them separately.
After $N$ realisations, one gets $N_L$ left-realisations labelled $\psi^{(l)}_L(t,\textbf{x})$ with $l=1,2,3,...,N_L$, and $N_R$ right-realisations labelled $\psi^{(r)}_R(t,\textbf{x})$ with $r=1,2,3,...,N_R$,
with $N_L+N_R=N$ and $N_L \approx N_R \approx N/2$.
Then, calculate two separate averages
\begin{align}
 \label{eq:ensembleAveChiralL}
 &\langle \psi_L(t,\textbf{x}) \rangle \equiv  \langle \psi_L \rangle (t,\textbf{x}) \equiv \Psi_L(t,\textbf{x}) := \lim \limits_{N \rightarrow \infty}
  \frac{1}{N_L} \sum \limits_{l=1}^{N_L}  \psi^{(l)}_L(t,\textbf{x})  \ , \qquad (t,\textbf{x}) \in \Upomega \\ 
   \label{eq:ensembleAveChiralR}
  &\langle \psi_R(t,\textbf{x}) \rangle \equiv  \langle \psi_R \rangle (t,\textbf{x}) \equiv \Psi_R(t,\textbf{x}) := \lim \limits_{N \rightarrow \infty}
  \frac{1}{N_R} \sum \limits_{r=1}^{N_R}  \psi^{(r)}_R(t,\textbf{x})  \ , \qquad (t,\textbf{x}) \in \Upomega 
\end{align}
whereby the series are assumed absolutely convergent.
Both ensemble averages, $\Psi_L$ and $\Psi_R$, must be taken into account in coming mathematical operations, for none is above the other nor can replace the other. 
Arguably, the consideration of chiral flows can also be extended to non-equally probable realisations, even with more than two possible outcomes
\footnote{For example, the chiral realisations might be: Left $\psi^{(n)}_L$, right $\psi^{(n)}_R$, up $\psi^{(n)}_U$, down $\psi^{(n)}_D \cdots$}. 
Such eventual unusual cases would correspond to bifurcations at fixed points in the phase space of the dynamical system.
To be of application, the formalism requires that $N_L/N_R \rightarrow k, \ 0<k\leq 1$ if $N_L<N_R$ (respectively, $N_R/N_L \rightarrow k, \ 0<k\leq 1$ if $N_R<N_L$), as $N\rightarrow\infty$.
\end{enumerate}
Henceforth, all derivations and explanations will be developed for Equiprobable Realisations, Eq. \eqref{eq:ensembleAve}, being the most frequent by far. 
The interested reader can try any demonstration with anyone of the other procedures.

An ensemble-averaged field $ \langle \psi(t,\textbf{x}) \rangle \equiv \Psi(t,\textbf{x})$ is also called a mean field.
Using the ensemble average defined above, it is possible to express mathematically the physical field $\psi \in \mathfrak{F}_{\Upomega}$ of any realisation as 
\begin{equation}
 \label{eq:ReynoldsDecomp}
 \psi^{(n)}(t,\textbf{x}) =  \Psi(t,\textbf{x}) + \psi'^{(n)}(t,\textbf{x})\ , \qquad (t,\textbf{x}) \in \Upomega 
\end{equation}
where $\psi'^{(n)}(t,\textbf{x})$, the so-called fluctuating component, is whatever that remains after subtracting the common field $ \Psi(t,\textbf{x})$ from each $\psi^{(n)}(t,\textbf{x})$.
According to Eq. \eqref{eq:ReynoldsDecomp}, the fluctuating components must have zero ensemble-average:
\begin{equation}
 \label{eq:fluctAve}
  \langle \psi'(t,\textbf{x}) \rangle = \lim \limits_{N \rightarrow \infty}
 \frac{1}{N} \sum \limits_{n=1}^{N}  \psi'^{(n)}(t,\textbf{x})=0 \ , \qquad (t,\textbf{x}) \in \Upomega 
\end{equation}
Note $\Psi(t,\textbf{x})$ is a mathematical entity, a construction of the mind and, in principle, might not be a physical field, 
i.e., something that could be witnessed in an experiment.
The field $\Psi(t,\textbf{x})$ belongs to a different functional space, called the {Mean Space} $\mathfrak{M}_{\Upomega}$, which is not the physical space $\mathfrak{F}_{\Upomega}$.
The mean space $\mathfrak{M}_{\Upomega}$ is a world of averages, in which one only finds ensemble-averaged quantities and fields (the mean space $\mathfrak{M}_{\Upomega}$ will be rigorously defined later).
In general, $\Psi(t,\textbf{x})$ will not be equal to any realisation $\psi^{(n)}(t,\textbf{x})$ and, possibly, not even near to it.
Furthermore, for given $\epsilon>0$, the mean field $\Psi(t,\textbf{x})$ does not depend on the chosen ensemble, always provided $N\rightarrow\infty$ and there is no systematic bias in the choice of realisations.
The peculiarities of any individual ensemble would vanish after division by $N$.

Note we are considering five different classes of functional spaces: The space $\mathfrak{F}_{\Upomega}$ of all physical flows in $\Upomega$,
the space of physical initial conditions of range $\epsilon$, $\psi_{0\langle \epsilon \rangle}(\textbf{x})$, 
the space of physical boundary conditions of range $\epsilon$, $\psi_{\partial \ulcorner \epsilon \lrcorner}(t,\textbf{s})$,
the ensemble of realisations stemming from those I\&BC of range $\epsilon$, $\mathfrak{E}_{\epsilon} \subset \mathfrak{F}_{\Upomega}$ (or simply $\mathfrak{E}$),
and the mean space $\mathfrak{M}_{\Upomega}$ that contains the ensemble-averaged (or mean) fields resulting from all possible ensembles $\mathfrak{E}$.
To each $\epsilon\mathop{>}0$ correspond different spaces $\psi_{0\langle \epsilon \rangle}(\textbf{x})$, $\psi_{\partial \ulcorner \epsilon \lrcorner}(t,\textbf{s})$ and $\mathfrak{E}$. 

The notion of ensemble-averaged or mean fields would not be complete without a study of how and when the particular features of each realisation are transmitted to them.
The mean fields that emerge from the ensemble of realisations would not be properly characterised without considering the influence of individual attributes of physical fields upon them.
A preliminary study about the conditions under which potential particularities of realisations affect the mean quantities and fields is conducted in \cref{sct:haps}.

\section{A definition of laminar and turbulent flow}
\label{sct:defLamTurb}

Before we engage into the formal definition of laminar and turbulent flow, it is sensible to examine what is considered conventional wisdom in this realm.
To avoid confusion, the notions the community has of these flows will be temporarily called `\textit{conventional laminar}' (or C-laminar) and `\textit{conventional turbulent}' (or C-turbulent), 
to distinguish them from the formal definitions that will be offered later.
Arguably, there exist general notions of what C-laminar and C-turbulent flows are, but a shared notion is not a formal definition.

Let us recall the beginning of \cref{sct:ensembAve}, with the set of repeated experiments and the concepts of realisation $\psi^{(n)}(t,\textbf{x})$ and ensemble average $\Psi(t,\textbf{x})$.
If the flow is kept C-laminar during all experiments and if we have been careful enough upon execution,
then conventional wisdom suggests that $\psi^{(n)}(t,\textbf{x}) \approx \psi^{(m)}(t,\textbf{x}), \ \forall \ n,m \in \mathbb{N}, \ \forall \ (t,\textbf{x}) \in \Upomega$
(with ``$\approx$'' loosely defined),
and this implies $\Psi(t,\textbf{x}) \approx \psi^{(n)}(t,\textbf{x}), \ \forall n \in \mathbb{N}, \ \forall (t,\textbf{x}) \in \Upomega$,
and the ensemble-averaged field would be alike to the physical field.
However, if the flow ends up being C-turbulent in each realisation, then we shall have $\Psi(t,\textbf{x}) \not \approx \psi^{(n)}(t,\textbf{x})$
regardless of how careful have we been in our experiments.
These notions, and a few others, should be implicit into any definition that might be proposed.

Let $\Upomega=[0,T_r] \times \Updelta \subset \mathbb{R}^4$ be the spacetime domain where a Newtonian incompressible flow develops, also called a physical flow.
The governing equations of this physical flow are the Navier-Stokes equation (NSE)
\begin{equation}
 \label{eq:NSE}
 \textbf{u}_{,0} + (\textbf{u}\cdot \nabla)\textbf{u} - \nu \nabla^2 \textbf{u} = -\rho^{-1} \ \nabla p  + \textbf{g}
\end{equation}
and the continuity equation
\begin{equation}
 \label{eq:continuity}
 \nabla \cdot \textbf{u} =0
\end{equation}
together with the initial condition $\textbf{u}(0,\textbf{x})\mathop{=}\textbf{u}_0(\textbf{x}), \ \textbf{x} \in \Updelta$, 
and $\textbf{u}_{\partial}(t,\textbf{s}), \ (t,\textbf{s}) \in [0,T_r]\times \partial \Updelta$, 
encompasses the boundary conditions for the velocity field or any of its derivatives.
The role of pressure in Eq. \eqref{eq:NSE} should be discussed.
A flow may be subject to a prescribed external pressure gradient, as in the case of a Hagen-Poiseuille flow, which is entirely defined by such a pressure gradient.
But in other cases, particularly in C-turbulent flow, the pressure field $p(t,\textbf{x})$ would take values that no longer respond to a prescribed pattern. 
For incompressible flow, the pressure field $p(t,\textbf{x})$ must obey the Poisson equation
\begin{equation}
 \label{eq:Poisson}
  \nabla^2 p = - \rho \nabla \cdot (\textbf{u} \cdot \nabla \textbf{u}) = - \rho \ u_{j,i}u_{i,j}
\end{equation}
whose solution is (see \citep[Eq. (11.3)]{Sob64})
\begin{equation}
\label{eq:PoissonSol}
 p(t,\textbf{x}')=\frac{\rho}{4 \pi}\int \limits_{\Updelta} \frac{\nabla \cdot (\textbf{u} \cdot \nabla \textbf{u})}{|\textbf{x}'-\textbf{x}|} \dd^3 \textbf{x} + \frac{1}{4 \pi} \int \limits_{\partial \Updelta} \left[ p(t,\textbf{s}) \frac{\partial}{\partial n} \left (\frac{1}{|\textbf{x}'-\textbf{s}|} \right ) 
 - \frac{1}{|\textbf{x}'-\textbf{s}|} \frac{\partial p(t,\textbf{s})}{\partial n} \right ]\dd \Upsigma \ , \quad \forall \ \textbf{x}' \in \Updelta
\end{equation}
The actual pressure field in the flow results from the combined action of the velocity field plus the externally prescribed pressure.
The solution \eqref{eq:PoissonSol} requires the previous knowledge of $p(t,\textbf{s})$ and $\partial p(t,\textbf{s})/\partial n$ at the boundary $\partial \Updelta$.
Therefore, these functions must be included in the boundary conditions of the problem, i.e., must be prescribed,
and they will be collectively denoted by $\uppi_{\partial}(t,\textbf{s})$.
In experiments, one can only aim to prescribe the boundary conditions $\uppi_{\partial}(t,\textbf{s})$; the actual pressure field $p(t,\textbf{x})$ existing in the flow depends also upon the velocity field $\textbf{u}(t,\textbf{x})$ through Eq. \eqref{eq:PoissonSol},
which blends the externally prescribed pressure with that created by the flow itself.
Note a most interesting fact: The source term in the Poisson Eq. \eqref{eq:Poisson} corresponds to the nonlinear term in the NSE \eqref{eq:NSE}.
Should NSE be linear, Eq. \eqref{eq:Poisson} would be the Laplace equation and the pressure field would be a harmonic function that depends exclusively on $\uppi_{\partial}(t,\textbf{s})$.
It is the nonlinearity that makes the pressure field behave so strangely.

Let $\mathfrak{F}_{\Upomega} \mathfrak{\subset} H^2(\Upomega)$ be the space of all possible physical flows associated to $\Upomega$, that is, 
the space of fields that are solutions of the governing equations for any flow that could be developed in $\Upomega$.
The space $\mathfrak{F}_{\Upomega}$ contains all flows that may be generated from all possible I\&BC compatible with $\Upomega$.
An element of $\mathfrak{F}_{\Upomega}$ is called a physical flow, denoted by $\mathfrak{f}_{\Upomega}(\textbf{u}_0,\textbf{u}_{\partial},\uppi_{\partial};\textbf{u},\boldsymbol{\pi})$, 
where $\textbf{u}_0(\textbf{x})$ is the initial velocity field, $\textbf{u}_{\partial}(t,\textbf{s})$ encompasses the boundary conditions for the velocity field, 
$\uppi_{\partial}(t,\textbf{s})$ are the boundary conditions for the pressure field according to Eq.  \eqref{eq:PoissonSol}, 
and $\textbf{u}(t,\textbf{x})$ and $ \bm{\pi}(t,\textbf{x})=\rho^{-1} \ \nabla p$ are the velocity and pressure-gradient fields that actually define the flow.
It is not necessary to prescribe an initial pressure field $p_0(\textbf{x})$, since it is uniquely derived from $\textbf{u}_0(\textbf{x})$ and $\uppi_{\partial}(0,\textbf{s})$ through Eq. \eqref{eq:PoissonSol}.
In general, $\mathfrak{F}_{\Upomega}$ itself is not a vector subspace of $H^2(\Upomega)$, since the linear combination of several physical flows is not a physical flow.
However, $\mathfrak{F}_{\Upomega}$ is a differentiable manifold whose local structure is briefly explored in \cref{sct:localStruct}.

Consider now the physical initial conditions of range
\footnote{Again, we could consider $\epsilon_1$ for $\textbf{u}_0$, $\epsilon_2$ for $\textbf{u}_{\partial}$ and $\epsilon_3$ for $\uppi_{\partial}$,
but such a choice would complicate matters with no significant contribution.}
$\epsilon$, $\textbf{u}_{0_{\langle \epsilon \rangle}}$, and the physical boundary conditions of range $\epsilon$, $\textbf{u}_{\partial_{ \ulcorner \epsilon \lrcorner}}$ and $\uppi_{\partial_{ \ulcorner \epsilon \lrcorner}}$,
as introduced in Defs. \ref{def:physInitCond} \& \ref{def:physBoundCond}, respectively.
Using different elements of these sets of physical I\&BC, an ensemble of realisations $\mathfrak{E}\mathop{\subset}\mathfrak{F}_{\Upomega}$ can be generated.
In this research, the sentences ``\textit{executing an experiment with I\&BC $\{\textbf{u}_0;\textbf{u}_{\partial},\uppi_{\partial}\}$}'', 
``\textit{running a computer simulation with I\&BC $\{\textbf{u}_0;\textbf{u}_{\partial},\uppi_{\partial}\}$}'' and
``\textit{solving Eq. \eqref{eq:NSE} with I\&BC $\{\textbf{u}_0;\textbf{u}_{\partial},\uppi_{\partial}\}$}'' are considered equivalent.
Therefore, realisations could be obtained by any of these three methods.
Note how the existence of $\uppi_{\partial_{ \ulcorner \epsilon \lrcorner}}$ introduces additional variability into the ensemble,
respect to one built exclusively from the velocity field conditions $\textbf{u}_0$ and $\textbf{u}_{\partial}$.

The ensemble $\mathfrak{E}$ thus created permits to define the corresponding mean fields $\textbf{U}_0(\textbf{x})$, $\textbf{U}_{\partial}(t,\textbf{s})$, 
$\Uppi_{\partial}(t,\textbf{s})$, $\textbf{U}(t,\textbf{x})$ and $ \bm{\Pi}(t,\textbf{x})=\rho^{-1} \ \nabla P$,
through any of the procedures specified by Eqs. \eqref{eq:ensembleAve}-\eqref{eq:ensembleAveChiralR}.
In turn, the physical fields can be expressed from the mean fields through the Reynolds decomposition of Eq. \eqref{eq:ReynoldsDecomp}.
Introducing the Reynolds decomposition into Eqs. \eqref{eq:NSE} \& \eqref{eq:continuity}, the Reynolds-averaged Navier-Stokes equation (RANSE) and the Reynolds-averaged continuity equation are obtained as
\begin{equation}
 \label{eq:RANSE}
 \textbf{U}_{,0} + (\textbf{U}\cdot \nabla)\textbf{U} - \nu \nabla^2 \textbf{U} = -\bm{\Pi} + \textbf{g} + \nabla \cdot \Re
\end{equation}
\begin{equation}
 \label{eq:Rcontinuity}
 \nabla \cdot \textbf{U}=\nabla \cdot \textbf{u}'^{(n)} =0 \ , \quad n \in \mathbb{N}
\end{equation}
with
\begin{equation}
\label{eq:ensAvg}
 \textbf{U}(t,\textbf{x})= \lim \limits_{N \rightarrow \infty}
 \frac{1}{N} \sum \limits_{n=1}^{N} \textbf{u} ^{(n)}(t,\textbf{x}) \ , \quad 
 \bm{\Pi}(t,\textbf{x})= \lim \limits_{N \rightarrow \infty} \frac{1}{N} \sum \limits_{n=1}^{N} \bm{\pi} ^{(n)}(t,\textbf{x}) \ , \qquad (t,\textbf{x}) \in \Upomega 
\end{equation}
\begin{equation}
 \textbf{u}^{(n)}(t,\textbf{x}) =  \textbf{U}(t,\textbf{x})+ \textbf{u}'^{(n)}(t,\textbf{x}) \ , \quad  \bm{\pi}^{(n)}(t,\textbf{x})=\bm{\Pi}(t,\textbf{x})+\bm{\pi}'^{(n)}(t,\textbf{x})\ , \qquad (t,\textbf{x}) \in \Upomega \ , \ n \in \mathbb{N}
\end{equation}
and where $\Re = \{\Re_{ij} \}$ is the Reynolds-stress tensor, defined by  
\begin{equation}
 \label{eq:RSS}
 \Re_{ij}(t,\textbf{x})= -\langle u'_i u'_j \rangle (t,\textbf{x}) = -\lim \limits_{N \rightarrow \infty}
 \frac{1}{N} \sum \limits_{n=1}^{N} {u'}_i ^{(n)}(t,\textbf{x}) \, {u'}_j ^{(n)}(t,\textbf{x}) \ , \qquad (t,\textbf{x}) \in \Upomega 
\end{equation}
with $N$ the number of realisations in the ensemble (it is assumed that $\textbf{G}=\textbf{g}$, i.e., the gravity field does not change between realisations).
The last term in Eq. \eqref{eq:RANSE} is the mean vector $\nabla \cdot \Re = \{\Re_{ij,i} \}= \{\Re_{ij,j} \}$.
Since the mean-velocity field $\textbf{U}(t,\textbf{x})$ is solenoidal, Eq. \eqref{eq:Rcontinuity}, the mean pressure  $P(t,\textbf{x})$ also obeys a Poisson equation, which takes the form:
\begin{equation}
 \label{eq:PoissonMean}
  \nabla^2 P = - \rho \nabla \cdot \left (\textbf{U} \cdot \nabla \textbf{U} - \nabla \cdot \mathfrak{R} \right) = - \rho \left ( U_{j,i} U_{i,j}- \mathfrak{R}_{ij,ij} \right)
\end{equation}
whose solution likewise is
\begin{equation}
\label{eq:PoissonSolMean}
 P(t,\textbf{x}')=\frac{\rho}{4 \pi}\int \limits_{\Updelta} \frac{\nabla \cdot (\textbf{U} \cdot \nabla \textbf{U}-\nabla \cdot \mathfrak{R})}{|\textbf{x}'-\textbf{x}|} \dd^3 \textbf{x} + \frac{1}{4 \pi} \int \limits_{\partial \Updelta} \left[ P(t,\textbf{s}) \frac{\partial}{\partial n} \left (\frac{1}{|\textbf{x}'-\textbf{s}|} \right ) 
 - \frac{1}{|\textbf{x}'-\textbf{s}|} \frac{\partial P(t,\textbf{s})}{\partial n} \right ]\dd \Upsigma \ , \quad \forall \ \textbf{x}' \in \Updelta
\end{equation}
Again, the solution \eqref{eq:PoissonSolMean} requires the previous knowledge of the boundary conditions $ P(t,\textbf{s})$ and $ \partial P(t,\textbf{s})/\partial n$, collectively denoted by $\Uppi_{\partial}(t,\textbf{s})$,
which coincide with the ensemble average of the boundary conditions $\uppi_{\partial_{ \ulcorner \epsilon \lrcorner}}$ that generate the ensemble of realisations.
And also, the source term in the Poisson Eq. \eqref{eq:PoissonMean} corresponds to the nonlinear terms in the RANSE \eqref{eq:RANSE}.
Should the RANSE be linear, Eq. \eqref{eq:PoissonMean} would be the Laplace equation and the mean-pressure field would be a harmonic function that depends exclusively on $\Uppi_{\partial}(t,\textbf{s})$.
It is the nonlinearity that makes the mean-pressure field behave so oddly.

In summary, the situation is as follows: On one hand we have physical flows (realisations) governed by NSE \eqref{eq:NSE}, and on the other mean flows governed by the RANSE \eqref{eq:RANSE}.
Except for the extra term $\nabla \mathop{\cdot} \Re$, which stems from the advective term $(\textbf{u} \mathop{\cdot} \nabla)\textbf{u}$ in \eqref{eq:NSE}, both equations are formally identical.
The term $\nabla \mathop{\cdot} \Re$ embodies one part of the nonlinearities of the NSE (the other part is contained in $(\textbf{U} \mathop{\cdot} \nabla)\textbf{U}$)
and introduces an additional mathematical difficulty in the RANSE, 
which is usually compensated with a reduction in the number of components or with a dependence upon fewer spacetime variables.
Mean flows are obtained through an ensemble-average procedure, i.e., a mathematical operation not dictated by any natural requisite.
In general, mean flows are not physical flows.

Mean flows can also be arranged into a functional space.
Let $\mathfrak{M}_{\Upomega} \mathop{\subset} H^2(\Upomega)$ be the space of mean flows, i.e., of solutions of the RANSE \eqref{eq:RANSE} \& \eqref{eq:Rcontinuity} compatible with the domain $\Upomega$.
An element of $\mathfrak{M}_{\Upomega}$ is a mean flow in $\Upomega$, denoted by $\mathfrak{m}^{\epsilon}_{\Upomega}(\textbf{U}_0, \textbf{U}_{\partial}, \Uppi_{\partial}; \textbf{U},\bm{\Pi}, \Re)$.
In general, $\mathfrak{M}_{\Upomega}$ is not a vector subspace of $H^2(\Upomega)$, though it is a differentiable manifold. 
Note there exists a direct relationship between the set of classes $\{ \textbf{u}_{0_{\langle \epsilon \rangle}}; 
\textbf{u}_{\partial_{ \ulcorner \epsilon \lrcorner}},\uppi_{\partial_{ \ulcorner \epsilon \lrcorner}}\}$ and the mean flow 
$\mathfrak{m}^{\epsilon}_{\Upomega}(\textbf{U}_0, \textbf{U}_{\partial}, \Uppi_{\partial}; \textbf{U},\bm{\Pi}, \Re)$ it generates and,
for given I\&BC $\{ \textbf{u}_{0}; \textbf{u}_{\partial},\uppi_{\partial}\}$, a different mean flow is obtained from each $\epsilon$.
However, the relationship between any realisation and its corresponding mean flow is extraordinarily weak, since many (formally infinite) realisations are needed to produce a mean flow.
The assumption that what is observed in a few realisations should also occur in the mean flow may be false.
And also, particularly in unsteady C-turbulent mean flow, it is highly unlikely that the Reynolds stress field $\Re(t,\textbf{x})$ be a function of the mean velocity field $\textbf{U}(t,\textbf{x})$,
at least not a \textit{conventional} function of mathematical analysis, as it is frequently assumed in CFD simulations.

It is an experimental fact that $\textbf{u}^{(n)}(t,\textbf{x}) \approx \textbf{u}^{(m)}(t,\textbf{x})$ for any two realisations $n$ and $m$ of a C-laminar flow,
whereas in C-turbulent flow the realisations could differ significantly.
This natural truth should be enforced into the mathematical definition of flow.
Using the notions of \cref{sct:physFunc} for C-laminar flows, the following postulate is acknowledged as true:  
\begin{postulate}[Realisations of C-laminar flow]
\label{pos:laminar}
 Let $\mathfrak{F}_{\Upomega}\mathop{\subset}H^2(\Upomega)$ be the functional space of physical flows in $\Upomega=[0,T_r] \times \Updelta \subset \mathbb{R}^4$.
 Let $\mathfrak{f}_{\Upomega}(\textbf{u}_0,\textbf{u}_{\partial},\uppi_{\partial};\textbf{u},\boldsymbol{\pi}) \mathop{\in} \mathfrak{F}_{\Upomega}$ be a physical C-laminar flow.
 For each $\epsilon\mathop{>}0$ there exists $\delta\mathop{>}0$ such that the set of physical I\&BC of range $\delta$, 
 $\{ \textbf{u}_{0_{\langle \delta \rangle}};  
\textbf{u}_{\partial_{ \ulcorner \delta \lrcorner}},\uppi_{\partial_{ \ulcorner \delta \lrcorner}} \}$, yields an ensemble of realisations $\mathfrak{E}_{\delta}$ satisfying 
 $\textbf{u}^{(n)}\cong_{\epsilon} \textbf{u}^{(m)}$ and  $\bm{\uppi}^{(n)}\cong_{\epsilon} \bm{\uppi}^{(m)}$ for any two realisations $n$ and $m$.
\end{postulate}
In case of C-turbulent flow, each realisation would be almost equal to any other only during a brief lapse of time at the very beginning, i.e.,
$\textbf{u}^{(n)}(t,\textbf{x})\approx \textbf{u}^{(m)}(t,\textbf{x})$ for $t \approx 0$.
As time advances, each realisation departs significantly from any other, taking every available branch in the infinite evolutionary paths of chaotic dynamical systems.
Such notions were already suggested at the beginning of this section and it seems appropriate to state them now in rigorous form.
These ideas may be embodied in another postulate:
\begin{postulate}[Beginning of realisations of C-turbulent flow]
\label{pos:turbulent}
 Let $\mathfrak{F}_{\Upomega}\mathop{\subset}H^2(\Upomega)$ be the functional space of physical flows in $\Upomega=[0,T_r] \times \Updelta \subset \mathbb{R}^4$.
 Let $\mathfrak{f}_{\Upomega}(\textbf{u}_0,\textbf{u}_{\partial},\uppi_{\partial};\textbf{u},\boldsymbol{\pi}) \mathop{\in} \mathfrak{F}_{\Upomega}$ be a physical C-turbulent flow.
 For each $\epsilon\mathop{>}0$ there exist $\delta\mathop{>}0$ and $\tau\mathop{>}0$ such that the set of physical I\&BC of range $\delta$, 
 $\{ \textbf{u}_{0_{\langle \delta \rangle}};  
\textbf{u}_{\partial_{ \ulcorner \delta \lrcorner}},\uppi_{\partial_{ \ulcorner \delta \lrcorner}} \}$, yields an ensemble of realisations $\mathfrak{E}_{\delta}$ satisfying  
 $\textbf{u}^{(n)}\stackrel{_\tau}{\cong}_{\epsilon} \textbf{u}^{(m)}$ and  $\bm{\uppi}^{(n)}\stackrel{_\tau} \cong_{\epsilon} \bm{\uppi}^{(m)}$ for any two realisations $n$ and $m$,
 where ``$f \stackrel{_\tau}{\cong}_{\epsilon} g$'' stands for 
\begin{equation}
 \label{eq:almostEqualT}
f \stackrel{_\tau}{\cong}_{\epsilon} g \Leftrightarrow \left \{ \int \limits_{0}^{\tau} \int \limits_{\Updelta} \left[ |f(t,\textbf{x})-g(t,\textbf{x})|^2 + |f_{,\alpha}(t,\textbf{x})-g_{,\alpha}(t,\textbf{x})|^2 +
|f_{,\alpha\beta}(t,\textbf{x})-g_{,\alpha\beta}(t,\textbf{x})|^2 \right] \dd t \dd \textbf{x}^3 \right \}^{\tfrac12} < \epsilon 
\end{equation}
 instead of the integral over $\Upomega=[0,T_r] \times \Updelta$.
\end{postulate}
Regrettably, neither Postulate \ref{pos:laminar} nor \ref{pos:turbulent} can offer a general relationship between $\epsilon$ and $\delta$, and $\tau$, 
which should be determined in each case.
It follows from Postulate \ref{pos:turbulent} that $\textbf{U}(t,\textbf{x}) \mathop{\approx} \textbf{u}^{(n)}(t,\textbf{x})$, $\forall \ n\in\mathbb{N}, \ 0 \leq t<\tau , \ \textbf{x} \in \Updelta$.
Generally, $\tau \mathop{\ll} T_r$ and the quasi-equality holds only for a short time $\tau$, which is related to the Lyapunov time of the flow considered as a dynamical system.
The fluctuating component $\textbf{u}'^{(n)}(t,\textbf{x})$ will not be negligible for $t > \tau$.
In any case, C-laminar or C-turbulent, it is $\langle\textbf{u}'(t,\textbf{x})\rangle\mathop{=}0, \ \forall \ (t,\textbf{x}) \mathop{\in} \Upomega$.

The reader may object that we have been discussing laminar and turbulent flows thus far, without providing a definition of what is actually meant with these terms.
Instead, the discussion has been based upon the general ideas already available in the pool of conventional wisdom. 
This rightful objection will be attended to now.
The fundamental idea, which leads to the formal definition of laminar and turbulent flow, is the following: In general, the mean fields $\textbf{U}(t,\textbf{x})$ and $\bm{\Pi}(t,\textbf{x})$ 
are not solution of NSE \eqref{eq:NSE}; and \textit{vice versa}, the physical fields $\textbf{u}(t,\textbf{x})$ and $\bm{\pi}(t,\textbf{x})$ are not solution of the RANSE \eqref{eq:RANSE}.
With spatially-changing Reynolds stresses, $ \nabla \cdot \Re\neq0$, there is no way to combine $\textbf{U}(t,\textbf{x})$ and $\bm{\Pi}(t,\textbf{x})$ in an equation like \eqref{eq:NSE}.
The definition of laminar flow is accomplished through a process of construction, expressed as follows
\footnote{As remarked before, it is possible to use $\delta_1$ and $\delta_2$ instead of a single $\delta$ in the definition,
but it is not an essential contribution.}:
\begin{definition}[Laminar flow]
 \label{def:laminarFlow}
Let $\mathfrak{F}_{\Upomega}\mathop{\subset}H^2(\Upomega)$ be the functional space of physical flows in $\Upomega\mathop{=}[0,T_r] \mathop{\times} \Updelta \mathop{\subset} \mathbb{R}^4$.
 Let $\mathfrak{f}_{\Upomega}(\textbf{u}_0,\textbf{u}_{\partial},\uppi_{\partial};\textbf{u},\boldsymbol{\pi}) \mathop{\in} \mathfrak{F}_{\Upomega}$ be a physical flow.
 Let $\mathfrak{E}_{\epsilon}$ be the ensemble of realisations generated from the physical I\&BC of range $\epsilon$, $\{\textbf{u}_{0_{\langle \epsilon \rangle}}; \textbf{u}_{\partial_{ \ulcorner \epsilon \lrcorner}},\uppi_{\partial_{ \ulcorner \epsilon \lrcorner}}\}$.
 Let $\mathfrak{m}^{\epsilon}_{\Upomega}(\textbf{U}_0, \textbf{U}_{\partial}, \Uppi_{\partial}; \textbf{U},\bm{\Pi}, \Re) \mathop{\in} \mathfrak{M}_{\Upomega}$ be the mean flow obtained from the ensemble $\mathfrak{E}_{\epsilon}$. 
 The physical flow $\mathfrak{f}_{\Upomega}(\textbf{u}_0,\textbf{u}_{\partial},\uppi_{\partial};\textbf{u},\boldsymbol{\pi})$ is laminar iff for each $\delta>0$ there exists $\epsilon_0 >0$ such that for all $\epsilon$,
 $0<\epsilon<\epsilon_0$, there exists a physical flow $\mathfrak{f}_{\Upomega}(\mathring{\textbf{u}}_0,\mathring{\textbf{u}}_{\partial},\mathring{\uppi}_{\partial};\mathring{\textbf{u}},\mathring{\boldsymbol{\pi}})\mathop{\in} \mathfrak{E}_{\epsilon}$ 
satisfying ${\textbf{U}} \mathop{\cong_{\delta}} \mathring{\textbf{u}}$ and $\bm{\Pi}\mathop{\cong_{\delta}} \mathring{\bm{\pi}}$. 
\end{definition}
The mean flow $\mathfrak{m}^{\epsilon}_{\Upomega}(\textbf{U}_0, \textbf{U}_{\partial}, \Uppi_{\partial}; \textbf{U},\bm{\Pi}, \Re)$ is arbitrarily proximate to an actual realisation
$\mathfrak{f}_{\Upomega}(\mathring{\textbf{u}}_0,\mathring{\textbf{u}}_{\partial},\mathring{\uppi}_{\partial};\mathring{\textbf{u}},\mathring{\boldsymbol{\pi}})\mathop{\in} \mathfrak{E}_{\epsilon}$,
i.e., to all practical effects, the mean flow is indistinguishable from a physical flow, from a solution of NSE.
In this case, the mean flow will be called a \textbf{quasi-solution} of NSE \eqref{eq:NSE}, a name used for a flow that is arbitrarily proximate to an actual solution.
One might expect to ever witness the mean flow in an experiment, within measuring instruments' uncertainty.
Postulate \ref{pos:laminar} could now be enunciated using ``laminar flow'' instead of ``C-laminar flow'', which has thus become unnecessary.
The above definition has the following mathematical consequence:
\begin{theorem}
\label{thr:laminarCond}
 Let $\mathfrak{f}_{\Upomega}(\textbf{u}_0,\textbf{u}_{\partial},\uppi_{\partial};\textbf{u},\boldsymbol{\pi})\in \mathfrak{F}_{\Upomega}$
 be a physical laminar flow, in accord with the formulation of Def. \ref{def:laminarFlow}.
 Then, the following approximate equality holds in the ensemble $\mathfrak{E}_{\epsilon}$
 \begin{equation}
  \label{eq:laminarCond}
  \lim \limits_{N\rightarrow \infty} \frac{N-1}{N^2} \sum \limits_{n=1}^{N}\left(\textbf{u}^{(n)}\cdot \nabla \right)\textbf{u}^{(n)} \approx_{ _\delta} 
  \lim \limits_{N\rightarrow \infty} \frac{1}{N^2} \sum \limits_{n=1}^{N} \sum \limits_{\substack{m=1 \\ m\neq n}}^{N} \left(\textbf{u}^{(n)}\cdot \nabla \right)\textbf{u}^{(m)}
 \end{equation}
where ``$f\approx_{ _\delta}g$'' stands for the canonical approximation (norm) in Lebesgue space $L^2(\Upomega)$
\begin{equation}
f\approx_{ _\delta}g \Leftrightarrow \left \{ \int \limits_{\Upomega} |f(t,\textbf{x})-g(t,\textbf{x})|^2  \dd t \dd \textbf{x}^3 \right \}^{\tfrac12} < \delta
\end{equation}
\end{theorem}
\begin{proof}
 If ${\textbf{U}} \mathop{\cong_{\delta}} \mathring{\textbf{u}}$ and $\bm{\Pi}\mathop{\cong_{\delta}} \mathring{\bm{\pi}}$,
 with $\{\mathring{\textbf{u}},\mathring{\bm{\pi}} \}$ a solution of the NSE \eqref{eq:NSE}, then it must be
 \begin{equation*}
   \textbf{U}_{,0} + (\textbf{U}\cdot \nabla)\textbf{U} - \nu \nabla^2 \textbf{U} + \bm{\Pi} - \textbf{g} \approx_{\delta}0
 \end{equation*}
 or, using the expansion in realisations of the mean fields (for the moment, $\lim_{N\rightarrow \infty}$ is omitted)
\begin{equation}
\label{eq:NSEdecomp1}
 \sum \limits_{n=1}^N \frac{\textbf{u}_{,0}^{(n)}}{N}+\left( \sum \limits_{n=1}^N \frac{\textbf{u}^{(n)} \cdot \nabla}{N} \right) \sum \limits_{m=1}^N \frac{\textbf{u}^{(m)}}{N}-
 \nu \sum \limits_{n=1}^N \frac{\nabla^2\textbf{u}^{(n)}}{N} + \sum \limits_{n=1}^N \frac{\bm{\pi}^{(n)}}{N} - \textbf{g} \approx_{\delta}0
\end{equation}

Since the single nonlinear term is $(\textbf{U}\cdot \nabla)\textbf{U}$, it suffices to consider only this term, which can be expressed as
\begin{equation*}
 \left( \sum \limits_{n=1}^N \frac{\textbf{u}^{(n)} \cdot \nabla}{N} \right) \sum \limits_{m=1}^N \frac{\textbf{u}^{(m)}}{N}= 
 \sum \limits_{n=1}^N \left( \textbf{u}^{(n)} \cdot \nabla \right) \frac{\textbf{u}^{(n)}}{N} + \frac{1-N}{N^2}\sum \limits_{n=1}^N \left( \textbf{u}^{(n)} \cdot \nabla \right) \textbf{u}^{(n)}+
 \frac{1}{N^2} \sum \limits_{n=1}^{N} \sum \limits_{\substack{m=1 \\ m\neq n}}^{N} \left(\textbf{u}^{(n)}\cdot \nabla \right)\textbf{u}^{(m)}
\end{equation*}
Substituting into Eq. \eqref{eq:NSEdecomp1} and rearranging the terms:
\begin{align}
\label{eq:NSEdecomp2}
 &\sum \limits_{n=1}^N \frac{\textbf{u}_{,0}^{(n)}}{N}+\sum \limits_{n=1}^N \left( \textbf{u}^{(n)} \cdot \nabla \right) \frac{\textbf{u}^{(n)}}{N}-
 \nu \sum \limits_{n=1}^N \frac{\nabla^2\textbf{u}^{(n)}}{N} + \sum \limits_{n=1}^N \frac{\bm{\pi}^{(n)}}{N} - \sum \limits_{n=1}^N \frac{\textbf{g}}{N} = \notag \\
 &\frac{1}{N}\sum \limits_{n=1}^N \left[ {\textbf{u}_{,0}^{(n)}}+\left( \textbf{u}^{(n)} \cdot \nabla \right) {\textbf{u}^{(n)}}-
 \nu {\nabla^2\textbf{u}^{(n)}} + {\bm{\pi}^{(n)}} - {\textbf{g}} \right ]\approx_{\delta} \notag \\
& \frac{N-1}{N^2}\sum \limits_{n=1}^N \left( \textbf{u}^{(n)} \cdot \nabla \right) \textbf{u}^{(n)}-
 \frac{1}{N^2} \sum \limits_{n=1}^{N} \sum \limits_{\substack{m=1 \\ m\neq n}}^{N} \left(\textbf{u}^{(n)}\cdot \nabla \right)\textbf{u}^{(m)} \notag
\end{align}
All terms in the sum of the left-hand side are identically zero, because they are solutions of the NSE (realisations).
Taking the limit as $N\rightarrow \infty$ proves the theorem.
\end{proof}
Mathematically, there are infinite ways to accomplish Eq. \eqref{eq:laminarCond}, since the realisations  $\{{\textbf{u}^{(n)}},{\bm{\pi}^{(n)}} \}$ are completely arbitrary, randomly chosen from the I\&BC.
However, from a physical point of view, only two possibilities stand out:
\begin{enumerate}
 \item The nonlinear component of the NSE \eqref{eq:NSE} is negligibly small in almost all realisations, $\left( \textbf{u}^{(n)} \mathop{\cdot} \nabla \right) \textbf{u}^{(n)}\approx_{\delta}0 $, 
 i.e., the NSE for this flow is quasi-linear.
 Since the ensemble-averaged field, Eq. \eqref{eq:ensAvg}, is a linear combination of physical fields and the NSE is quasi-linear, 
 then a linear combination of solutions must be a quasi-solution of NSE \eqref{eq:NSE}.
Further, the pressure field would be a quasi-harmonic function of the space variables
\footnote{It is for this very reason, for example, that in a Hagen-Poiseuille flow the pressure gradient is constant and it is superfluous to consider the integral of Eq. \eqref{eq:PoissonSol}}.
If follows from these arguments that the RANSE \eqref{eq:RANSE} is also quasi-linear and the mean-pressure is quasi-harmonic. 
\item The nonlinear term of a realisation $n$ is the quasi-ensemble average of all other nonlinear terms of the remaining realisations in the ensemble $\mathfrak{E}_{\epsilon}$, i.e., 
the following two approximate equations hold for almost all realisations:
\begin{equation}
 \label{eq:condLaminarEq}
 \left( \textbf{u}^{(n)} \cdot \nabla \right) \textbf{u}^{(n)} \approx_{_\delta} \lim \limits_{N\rightarrow \infty}  \sum \limits_{\substack{m=1 \\ m\neq n}}^{N} \left( \textbf{u}^{(m)} \cdot \nabla \right) \frac{\textbf{u}^{(n)}}{N-1}   ,\quad 
  \left( \textbf{u}^{(n)} \cdot \nabla \right) \textbf{u}^{(n)} \approx_{_\delta} \lim \limits_{N\rightarrow \infty}  \frac{ \textbf{u}^{(n)} \cdot \nabla }{N-1} \sum \limits_{\substack{m=1 \\ m\neq n}}^{N}  \textbf{u}^{(m)}
\end{equation}
which would in turn be satisfied if a single sufficient condition were true for almost every realisation $n$:
\begin{equation}
 \label{eq:singleCondLaminarEq}
 \textbf{u}^{(n)} (t,\textbf{x}) \approx_{_\delta} \lim \limits_{N\rightarrow \infty} \frac{1}{N-1} \sum \limits_{\substack{m=1 \\ m\neq n}}^{N} \textbf{u}^{(m)} (t,\textbf{x}) 
\end{equation}
This possibility 2. means that the realisations are so close among them, 
that almost every nonlinear term can be expressed as the quasi-ensemble average of the others
(or, alternatively, almost all realisations are $\delta$-equal to the ensemble average of the remaining realisations, which is a sufficient condition).
The NSE may be nonlinear, but their solutions evolve within the non-chaotic region of the phase space and are thus quite repeatable.
And being repeatable is the cornerstone of the concept of laminar flow.
This peculiar type of nonlinearity will be called \textbf{restricted nonlinearity}.
Note all realisations in the ensemble are arbitrary, i.e., if this second possibility were true, then it would be true regardless of the actual realisations forging Eq. \eqref{eq:condLaminarEq}. 
\end{enumerate}

The reader has probably noticed the ``\textit{almost all}'' tag in the above remarks.
Not all realisations of the ensemble must fulfil conditions 1 or 2 above. 
It suffices that the number of exceptions to those rules be a lesser infinite than $N$ (see Def. \ref{def:lesserInf} in \cref{sct:haps}).
In summary, the realm of laminar flows encompasses the following three cases: (i) fully linear ($\left( \textbf{u}^{(n)} \mathop{\cdot} \nabla \right) \textbf{u}^{(n)} \mathop{=}0$),
(ii) quasi-linear ($\left( \textbf{u}^{(n)}\mathop{\cdot} \nabla \right) \textbf{u}^{(n)}\mathop{\approx_{\delta}}0$), and (iii) restricted nonlinearity (Eq. \eqref{eq:condLaminarEq}).
It follows that laminar flows preclude arbitrary nonlinearity and are only compatible with these three types.
In fact, the restricted nonlinearity condition (iii) is the most comprehensive and includes (i) and (ii) as special cases.

Furthermore, for $\epsilon$ sufficiently small, the notion of being laminar is actually a property of the ensemble $\mathfrak{E}_{\epsilon}$. 
All realisations satisfying Eq. \eqref{eq:singleCondLaminarEq} can be considered equivalent, and thus constitute an equivalence class made up of individual laminar physical flows.
Any realisation satisfying Eq. \eqref{eq:singleCondLaminarEq} is just a representative of its laminar equivalence class (denoted by an overline).
 
The definition of turbulent flow is likewise based upon a process of construction:
\begin{definition}[Turbulent flow]
 \label{def:turbulentFlow}
Let $\mathfrak{F}_{\Upomega}\mathop{\subset}H^2(\Upomega)$ be the functional space of physical flows in $\Upomega\mathop{=}[0,T_r] \mathop{\times} \Updelta \mathop{\subset} \mathbb{R}^4$.
Let $\mathfrak{f}_{\Upomega}(\textbf{u}_0,\textbf{u}_{\partial},\uppi_{\partial};\textbf{u},\boldsymbol{\pi}) \mathop{\in} \mathfrak{F}_{\Upomega}$ be a physical flow.
 Let $\mathfrak{E}_{\epsilon}$ be the ensemble of realisations $\{{\textbf{u}}^{(n)}, {\bm{\pi}^{(n)}}\}$ generated from the physical I\&BC of range $\epsilon$, $\{\textbf{u}_{0_{\langle \epsilon \rangle}}; \textbf{u}_{\partial_{ \ulcorner \epsilon \lrcorner}},\uppi_{\partial_{ \ulcorner \epsilon \lrcorner}}\}$.
Let $\mathfrak{m}^{\epsilon}_{\Upomega}(\textbf{U}_0, \textbf{U}_{\partial}, \Uppi_{\partial}; \textbf{U},\bm{\Pi}, \Re) \mathop{\in} \mathfrak{M}_{\Upomega}$ be the mean flow obtained from the ensemble $\mathfrak{E}_{\epsilon}$. 
The flow $\mathfrak{f}_{\Upomega}(\textbf{u}_0,\textbf{u}_{\partial},\uppi_{\partial};\textbf{u},\boldsymbol{\pi})$ is turbulent
iff for each $\epsilon \mathop{>}0$ there exists $M\mathop{>}0$ such that ${\textbf{U}} \mathop{\not \cong_{_M}} {\textbf{u}}^{(n)}$ , $\bm{\Pi}\mathop{\not\cong_{_M}} {\bm{\pi}^{(n)}}$
$\forall \ n \mathop{\in} \mathbb{N}$, for all realisations of the ensemble $\mathfrak{E}_{\epsilon}$, where ``$f \mathop{\not \cong_{_M}} g$'' stands for 
\begin{equation}
 \label{eq:almostEqualOmega}
f \mathop{\not \cong_{_M}} g \Leftrightarrow M\mathop{<}\left \{ \int \limits_{\Upomega} \left[ |f(t,\textbf{x})-g(t,\textbf{x})|^2 + |f_{,\alpha}(t,\textbf{x})-g_{,\alpha}(t,\textbf{x})|^2 +  |f_{,\alpha\beta}(t,\textbf{x})-g_{,\alpha\beta}(t,\textbf{x})|^2 \right] \dd t \dd \textbf{x}^3 \right \}^{\tfrac12} \mathop{<}+\infty 
\end{equation}
\end{definition}
If the flow $\mathfrak{f}_{\Upomega}$ is turbulent then not a single experiment would ever yield as a result the velocity field $\textbf{U}(t,\textbf{x})$;
such a flow $\mathfrak{m}^{\epsilon}_{\Upomega}(\textbf{U}_0, \textbf{U}_{\partial}, \Uppi_{\partial}; \textbf{U},\bm{\Pi}, \Re)$
cannot occur in nature, it would be a non-natural flow.
Not a single realisation in $\mathfrak{E}_{\epsilon}$ would be proximate to $\mathfrak{m}^{\epsilon}_{\Upomega}(\textbf{U}_0, \textbf{U}_{\partial}, \Uppi_{\partial}; \textbf{U},\bm{\Pi}, \Re)$.
Mathematically, the nonlinearity of the NSE \eqref{eq:NSE} is significant and cannot be ignored, the solutions are well within the chaotic domain of the phase space,
and a mean field (which is a linear combination of physical fields) cannot be anywhere near of any physical field.
Further, if time $T_r$ is sufficiently long, then $M$ does not depend on the chosen $\epsilon$.
Turbulent flows are thus a consequence of \textbf{general nonlinearity}, as opposed to the restricted nonlinearity proper of laminar flows.
Postulate \ref{pos:turbulent} holds and can be enunciated with ``turbulent flow'' instead of ``C-turbulent flow', an expression which is no longer necessary.
And obviously, by construction, $\textbf{U}(t,\textbf{x})$ and $\bm{\Pi}(t,\textbf{x})$ are always a solution of the RANSE \eqref{eq:RANSE}, 
regardless of the flow $\mathfrak{f}_{\Upomega}$ being laminar or turbulent.

Let us now prove another intuitive piece of common knowledge: If a given physical flow may be satisfactorily modelled with a sufficiently regular function, then the flow is laminar.
Physical turbulent flows would not be modelled with simple analytical functions.  

\begin{theorem}
\label{thr:analyticity}
 Let $\mathfrak{f}_{\Upomega}(\textbf{u}_0,\textbf{u}_{\partial},\uppi_{\partial};\textbf{u},\boldsymbol{\pi}) \mathop{\in} \mathfrak{F}_{\Upomega}$ be a physical flow,
characterised by explicit analytic functions $\textbf{u}(t,\textbf{x}), \boldsymbol{\pi}(t,\textbf{x}) \in C^{\infty}(\Upomega)$,
whose parameters stem from the I\&BC $\{\textbf{u}_0;\textbf{u}_{\partial},\uppi_{\partial}\}$, i.e., all partial derivatives exist and are continuous in $\Upomega$.
Then, the physical flow $\mathfrak{f}_{\Upomega}$ is laminar and it is a representative of its laminar equivalence class $\overline{\mathfrak{f}}_{\Upomega}$.
\end{theorem}
\begin{proof}
The proof begins by constructing the ensemble $\mathfrak{E}_{\epsilon}$ associated to the physical flow 
$\mathfrak{f}_{\Upomega}(\textbf{u}_0,\textbf{u}_{\partial},\uppi_{\partial};\textbf{u},\boldsymbol{\pi})$, for $\epsilon$ sufficiently small, i.e.,
the ensemble generated from $\{ \textbf{u}_{0_{\langle \epsilon \rangle}}; 
\textbf{u}_{\partial_{ \ulcorner \epsilon \lrcorner}},\uppi_{\partial_{ \ulcorner \epsilon \lrcorner}}\}$.
For the moment, it will be assumed that $\mathfrak{E}_{\epsilon}$ has a finite number $N$ of realisations.
Let $\{\textbf{u}^{(n)},\boldsymbol{\pi}^{(n)} \}$ be the $n^{th}$ realisation of the ensemble, 
and let $\psi^{(n)}(t,\textbf{x}) \mathop{\equiv} \psi^{(n)}(x_{\alpha})\mathop{\equiv}\psi^{(n)}(\undertilde{\textbf{\textit{x}}}) $ 
(see Sect. \ref{sct:nomencl}) be any component of $\{\textbf{u}^{(n)}(\undertilde{\textbf{\textit{x}}}),\boldsymbol{\pi}^{(n)}(\undertilde{\textbf{\textit{x}}}) \}$.
Each function $\psi^{(n)}(\undertilde{\textbf{\textit{x}}})$ is analytic by hypothesis.
To calculate $\psi^{(n)}(\undertilde{\textbf{\textit{x}}})$, select a fixed spacetime point $\undertilde{\mathring{\textbf{\textit{x}}}} \mathop{\equiv} (\mathring{x}_{\alpha}) \mathop{\equiv} (\mathring{t},\mathring{\textbf{x}}) \mathop{\in} \Upomega$ (the pivot point)
such that the rectilinear segment $[\undertilde{\textbf{\textit{x}}}-\mathring{\undertilde{\textbf{\textit{x}}}}]\mathop{=}\lambda\undertilde{\textbf{\textit{x}}}+(1-\lambda)\mathring{\undertilde{\textbf{\textit{x}}}}$, 
$\lambda \mathop{\in} [0,1]$, is entirely included in $\Upomega$, 
$[\undertilde{\textbf{\textit{x}}}-\mathring{\undertilde{\textbf{\textit{x}}}}] \mathop{\subset} \Upomega$.
The Taylor's theorem dictates that, for each $k \in \mathbb{N}$ and for every $\undertilde{\textbf{\textit{x}}}\in \Upomega$ compatible with the pivot point $\mathring{\undertilde{\textbf{\textit{x}}}}$,
the function $\psi^{(n)}(\undertilde{\textbf{\textit{x}}})$ may be expressed as (see \citep[Sec. 12.14]{Apo81}):
\begin{align}
 \label{eq:TaylorDec}
 &\psi^{(n)}(\undertilde{\textbf{\textit{x}}}) = \psi^{(n)}(\mathring{\undertilde{\textbf{\textit{x}}}}) +
 \partial_{\alpha} \psi^{(n)}(\mathring{\undertilde{\textbf{\textit{x}}}}) (x_{\alpha}-\mathring{x}_{\alpha})+
 \frac{\partial^2_{\alpha\beta} \psi^{(n)}(\mathring{\undertilde{\textbf{\textit{x}}}})}{2!} (x_{\alpha}-\mathring{x}_{\alpha})(x_{\beta}-\mathring{x}_{\beta}) +\cdots +\\
 &\frac{ \partial^{k-1}_{\alpha \beta \cdots \gamma} \psi^{(n)}(\mathring{\undertilde{\textbf{\textit{x}}}})}{(k-1)!} (x_{\alpha}-\mathring{x}_{\alpha})(x_{\beta}-\mathring{x}_{\beta})\cdots(x_{\gamma}-\mathring{x}_{\gamma})+
\frac{ \partial^{k}_{\alpha \beta \cdots \gamma\delta} \psi^{(n)}(\undertilde{\textbf{\textit{z}}}_n)}{k!} (x_{\alpha}-\mathring{x}_{\alpha})(x_{\beta}-\mathring{x}_{\beta})\cdots(x_{\gamma}-\mathring{x}_{\gamma})(x_{\delta}-\mathring{x}_{\delta}) \notag
\end{align}
where $\undertilde{\textbf{\textit{z}}}_n \mathop{\in} [\undertilde{\textbf{\textit{x}}}-\mathring{\undertilde{\textbf{\textit{x}}}}]$ 
is a spacetime point within the rectilinear segment, and the last term constitutes the remainder $r^{(n)}_k(\undertilde{\textbf{\textit{x}}})$, 
which tends to zero as $k\rightarrow \infty$.
Note $\psi^{(n)}(\undertilde{\textbf{\textit{x}}})$ is a multivariate polynomial of maximum degree $k$ in the four variables $(x_{\alpha})$,
since functions calculated at $\mathring{\undertilde{\textbf{\textit{x}}}}$ and $\undertilde{\textbf{\textit{z}}}_n$ are considered constant real numbers (coefficients of the polynomial).
Four-variate polynomials of maximum degree $k$ form a vector space of dimension $(k+4)!/k! 4!$.
Henceforth, all functions $\psi^{(n)}(\undertilde{\textbf{\textit{x}}})$ should be understood as four-variate polynomials of maximum degree $k$.

Consider now the $m^{th}$ realisation of the ensemble $\mathfrak{E}_{\epsilon}$, $\psi^{(m)}(\undertilde{\textbf{\textit{x}}})$.
By hypothesis, $\psi^{(m)}(\undertilde{\textbf{\textit{x}}})$ has the same functional form as $\psi^{(n)}(\undertilde{\textbf{\textit{x}}})$, since their I\&BC are infinitesimally proximate to each other
(e.g., if $\mathfrak{f}_{\Upomega}$ is a Hagen-Poiseuille flow, then both realisations are parabolas,
$\psi^{(n)}(\undertilde{\textbf{\textit{x}}}) \mathop{\equiv} u^{(n)}(r)\mathop{=}\mathop{-}\left(\Pi^{(n)}/4\mu\right)\left(1-r^2/R^2\right)$
and $\psi^{(m)}(\undertilde{\textbf{\textit{x}}}) \mathop{\equiv} u^{(m)}(r) \mathop{=}\mathop{-}\left(\Pi^{(m)}/4\mu\right)\left(1-r^2/R^2\right)$, with $\Pi^{(n)}\mathop{\approx}\Pi^{(m)}$).
It follows that the Taylor polynomial of $\psi^{(m)}(\undertilde{\textbf{\textit{x}}})$ is arbitrarily proximate to Eq. \eqref{eq:TaylorDec},
i.e., for each $0\leq j<k$ it is \( \partial^{j}_{\alpha \beta \cdots \gamma} \psi^{(m)}(\mathring{\undertilde{\textbf{\textit{x}}}}) \mathop{\approx} \partial^{j}_{\alpha \beta \cdots \gamma} \psi^{(n)}(\mathring{\undertilde{\textbf{\textit{x}}}}) \).
This will occur for any realisation of the ensemble $\mathfrak{E}_{\epsilon}$, provided $\epsilon$ is sufficiently small.
Since all realisations $\psi^{(n)}(\undertilde{\textbf{\textit{x}}})$ are four-variate polynomials of maximum degree $k$, i.e., are vectors of a vector space,
any linear combination of them is also a vector of the same vector space.
In particular, 
\[ \Psi_{(N)}(\undertilde{\textbf{\textit{x}}})  = \frac{1}{N} \sum \limits_{n=1}^{N}  \psi^{(n)}(\undertilde{\textbf{\textit{x}}})   \]
is also a four-variate polynomial of maximum degree $k$, similar to Eq. \eqref{eq:TaylorDec}.
It follows from linearity that the $j^{th}$, $0\leq j<k$, coefficient of $\Psi_{(N)}(\undertilde{\textbf{\textit{x}}})$ will be:
\[ \frac{ \partial^{j}_{\alpha \beta \cdots \gamma} \Psi_{(N)}(\mathring{\undertilde{\textbf{\textit{x}}}})}{j!} =   \frac{1}{N} \sum \limits_{n=1}^{N} \frac{ \partial^{j}_{\alpha \beta \cdots \gamma} \psi^{(n)}(\mathring{\undertilde{\textbf{\textit{x}}}})}{j!}  \]
with all terms of the sum being infinitesimally proximate to one another, since each $\psi^{(n)}(\undertilde{\textbf{\textit{x}}})$ is analytical.
In these conditions, it is verified that 
\(\partial^{j}_{\alpha \beta \cdots \gamma} \Psi_{(N)}(\mathring{\undertilde{\textbf{\textit{x}}}}) \approx \partial^{j}_{\alpha \beta \cdots \gamma} \psi^{(n)}(\mathring{\undertilde{\textbf{\textit{x}}}}) \),
 $0 \leq j <k$, for almost every $n$, $1 \leq n \leq N$.
But this statement is actually very close to the restricted nonlinearity condition \eqref{eq:singleCondLaminarEq}, which would prove that the physical flow $\mathfrak{f}_{\Upomega}$ is laminar,
because the mean flow is a quasi-solution of the NSE.
We are almost done, it only remains to check the remainders $r^{(n)}_k(\undertilde{\textbf{\textit{x}}})$ in all realisations.
In principle, there is no rule relating $\undertilde{\textbf{\textit{z}}}_n $
and $\undertilde{\textbf{\textit{z}}}_m $ for any two realisations $n$ and $m$.
However, this problem disappears upon doing $k \rightarrow \infty$, as $r^{(n)}_k(\undertilde{\textbf{\textit{x}}}) \rightarrow 0$ for all $n$.
Taking the limit with $N\rightarrow \infty$ proves the theorem. 
\end{proof}
In \cref{sct:KarmanVortex} there is an actual example applying this theorem, where the reader may examine the details of the proposed method.
Thus far, we have considered equivalence classes for the I\&BC, see \cref{sct:physFunc}.
Now we also introduce the notion of equivalence class for the resulting laminar flows.
If $\epsilon$ is sufficiently small, the realisations of the ensemble $\mathfrak{E}_{\epsilon}$ (and the mean flow itself) may be considered members of an equivalence class, i.e.,
they may be considered equivalent laminar flows for all practical purposes, indistinguishable from one another.
Turbulent flows cannot be considered equivalent in this sense; this is another feature that distinguishes laminar flows from turbulent ones.

An equivalent condition for laminar flow is offered next, whose demonstration is trivial and thus omitted.
\begin{lemma}
\label{thr:zeroRey}
 A physical flow $\mathfrak{f}_{\Upomega}(\textbf{u}_0,\textbf{u}_{\partial},\uppi_{\partial};\textbf{u},\boldsymbol{\pi}) \mathop{\in} \mathfrak{F}_{\Upomega}$
 is laminar iff for each $\delta\mathop{>}0$ there exists $\epsilon_0 \mathop{>}0$ such that for all $\epsilon$, $0<\epsilon<\epsilon_0$, 
 the mean flow $\mathfrak{m}^{\epsilon}_{\Upomega}(\textbf{U}_0, \textbf{U}_{\partial}, \Uppi_{\partial}; \textbf{U},\bm{\Pi}, \Re) \mathop{\in} \mathfrak{M}_{\Upomega}$ 
 generated from $\{ \textbf{u}_{0_{\langle \epsilon \rangle}}; 
\textbf{u}_{\partial_{ \ulcorner \epsilon \lrcorner}},\uppi_{\partial_{ \ulcorner \epsilon \lrcorner}}\}$
satisfies $\nabla \cdot \Re \mathop{ \cong_{\delta}}0$.
\end{lemma}
The restricted nonlinearity of laminar flows is incompatible with non-homogeneous Reynolds stresses.
Conversely, the physical flow $\mathfrak{f}_{\Upomega}(\textbf{u}_0,\textbf{u}_{\partial},\uppi_{\partial};\textbf{u},\boldsymbol{\pi})$ is turbulent iff for every $\epsilon \mathop{>}0$ there exists $M\mathop{>}0$,
not necessarily small, such that the mean flow $\mathfrak{m}^{\epsilon}_{\Upomega}(\textbf{U}_0, \textbf{U}_{\partial}, \Uppi_{\partial}; \textbf{U},\bm{\Pi}, \Re)$ 
satisfies $\nabla \cdot \Re \ \mathop{\not \cong_{_M}}  0$.
The general nonlinearity of turbulent flows brings forth the Reynolds stresses.

Note that a single-point statistics quantity such as $\Re(t,\textbf{x})$ is sufficient to determine whether the flow is laminar or turbulent;
there is no need to resort to two-point statistics for this task.
The abstract argumentation developed in \cref{sct:haps} about haps and general fields $\psi(t,\textbf{x})$, 
may now be repeated for the specific case of $\psi(t,\textbf{x})=u'_i(t,\textbf{x)}\, u'_j(t,\textbf{x})=u'_i u'_j$,
because the field
\footnote{Oddly enough, the field $u'_i u'_j$ can hardly be called a physical field, since it depends directly upon the mean field $\textbf{U}(t,\textbf{x})$, which is not a physical field.}
$u'_i u'_j$ emerges as the best candidate to assess whether a given flow is laminar or turbulent.
It makes sense to identify a hap's limits with the spacetime domain in which $|u'_i u'_j|>\epsilon'$, for given $\epsilon'>0$.
For instance, if $u'_iu'_j$ is sparse, even if it is significant in some realisations, then $\Re=0$ as demonstrated in Theor. \ref{thr:sparse} of \cref{sct:haps}. 
Thus, it might be possible (though very unlikely) to have seemingly turbulent realisations of a laminar flow, according to the definition introduced in this research.
Note the relationship between $\Re_{ij}(t,\textbf{x})$ and any of the components $u'^{(n)}_i(t,\textbf{x})$ is absolutely negligible, not to say infinitesimal.
The idea of extracting information about $\Re_{ij}(t,\textbf{x})$ from just a few realisations is somewhat commonplace, but utterly unfounded.
As it is the idea of obtaining $\Re_{ij}(t,\textbf{x})$ from the mean velocity field  $\textbf{U}(t,\textbf{x})$, since nothing of the sort can be observed in Eq. \eqref{eq:RSS}.
Recall also that the notion of laminar flow does not require $u'_i u'_j=0$, which is physically impossible, but rather $u'_i u'_j \cong_{\delta} 0$ in almost all realisations.

The Reynolds stresses belong exclusively to the mean space $\mathfrak{M}_{\Upomega}$ and are not physical fields.
Physical flows do not '\textit{feel}' the Reynolds stresses. 
A physical flow is exclusively driven by pressure gradient, gravity, other body forces and viscosity, all of them included in the NSE.
The forces due to viscosity could be moderate, as in a laminar flow, or could be quite intense, as in a turbulent flow.
But in either case, the physical flow undergoes the effect of individual viscous force patterns.
The Reynolds stresses only appear in $\mathfrak{M}_{\Upomega}$, after executing the averaging procedures.
Only mean flows are subject to Reynolds stresses; it is an exclusive feature of the mean space $\mathfrak{M}_{\Upomega}$.
There is no way that a physical flow could be influenced by the Reynolds stresses, because they only emerge after an averaging process.

It is important to realise that the proposed definition of laminar and turbulent flows is based upon the dynamic behaviour of the mean flow created from the ensemble.
It is a definition by construction.
Indeed, it is the different properties of the mean-velocity field that permits to distinguish between both types of flow.
A mean-velocity field $\textbf{U}(t,\textbf{x})$ that (almost) fulfils NSE implies that the whole ensemble of physical flows $\mathfrak{E}_{\epsilon}$ may be classified as laminar flows,
including $\textbf{U}(t,\textbf{x})$ itself (within instruments' uncertainty).
A mean-velocity field $\textbf{U}(t,\textbf{x})$ that is neatly different from \textbf{all} realisations $\textbf{u}^{(n)}(t,\textbf{x})$ of its ensemble $\mathfrak{E}_{\epsilon}$,
reveals that those realisations should be classified as turbulent flows.
The key concept is repeatability; if the realisations are repeatable (chiral flows notwithstanding) the flow qualifies as laminar, otherwise it ought to be considered turbulent.

The previous statements notwithstanding, it should be agreed that Defs. \ref{def:laminarFlow} \& \ref{def:turbulentFlow} would seem incomplete to anyone who has devoted time to study and ponder about turbulent flows.
Physicists and engineers alike would be missing, 
at least, two aspects in those definitions: (i) no relationship has been proposed between $\epsilon$ and the Reynolds stress tensor $\Re$ obtained from it;
and (ii) it is not obvious how to classify flows with small Reynolds-stress divergence, $\nabla \cdot \Re \approx 0$, e.g., 
those appearing in the grey-shaded region of the Moody chart, or those corresponding to transition to turbulence.
Regrettably, no answer can be offered to the first aspect; there exists an intuition that the smaller $\epsilon$, the smaller $\Re$,
although for fully turbulent flows the size of $\epsilon$ would seem irrelevant, given enough time.

Regarding the second aspect, the correct procedure would be to subdivide the flow's domain $\Upomega$ into disjoint subdomains $\Upomega_{\lambda}$,
with $\bigcup_{\lambda} \Upomega_{\lambda}=\Upomega$,
such that in each one the laminar or turbulent behaviour be obvious or significantly dominant (recall every $\Upomega_{\lambda}$ must include a time interval and a spatial extent).
It would then be appropriate to consider piecewise-laminar-turbulent flows.
Still, the characterisation of flows in transition to turbulence will always be problematic.
Probably, most of the examples that readers will think of to challenge our definitions would come from this grey-shaded region, 
which would include the full inventory of instabilities, critical points and bifurcations proper of chaotic systems.
However, the procedure of gathering an ensemble of realisations and calculating the ensemble-averaged fields is unambiguous, and will always produce mathematically exact results.
Whatever the results, the proposed definitions will always determine whether the physical flows should be classified as laminar or turbulent, with no ambiguity whatsoever.
In contrast, the conventional notions of C-laminar and C-turbulent flows would be scarcely applicable to the grey-shaded region, and would reduce the attempt to classify flows to a matter of opinion.
The proposed definitions provide a previously unavailable criterion.

\section{Local structure of the manifold of physical flows}
\label{sct:localStruct}
Readers acquainted with the stability theory of laminar flows will find the contents of this section somewhat familiar, 
although the approach taken here differs from that in classic texts, since the scope of the theory is also different.
Let $\mathfrak{f}_{\Upomega}(\textbf{u}_0,\textbf{u}_{\partial},\uppi_{\partial};\textbf{u},\boldsymbol{\pi})$ be a solution of NSE for a given flow in $\Upomega\subset \mathbb{R}^4$, 
with known I\&BC within the physical I\&BC of range $\epsilon$ 
$\{ \textbf{u}_{0_{\langle \epsilon \rangle}};  
\textbf{u}_{\partial_{ \ulcorner \epsilon \lrcorner}},\uppi_{\partial_{ \ulcorner \epsilon \lrcorner}}\}$.
Let $\mathfrak{f}_{\Upomega}(\textbf{v}_0,\textbf{v}_{\partial},\upeta_{\partial};\textbf{v},\boldsymbol{\eta})$ be another solution of NSE for the same type of flow, 
with different I\&BC also within the same equivalence class
$\{ \textbf{u}_{0_{\langle \epsilon \rangle}};  
\textbf{u}_{\partial_{ \ulcorner \epsilon \lrcorner}},\uppi_{\partial_{ \ulcorner \epsilon \lrcorner}}\}$.
Assume $\textbf{v}(t,\textbf{x})$ and $\bm{\eta}(t,\textbf{x})$ can be expressed as
\begin{equation}
\label{eq:localDesc}
 \textbf{v}(t,\textbf{x}) = \textbf{u}(t,\textbf{x}) + \bm{\delta} \textbf{u}(t,\textbf{x}),\quad \bm{\eta}(t,\textbf{x})= \bm{\pi}(t,\textbf{x}) + \bm{\delta \pi}(t,\textbf{x})
\end{equation}
for some functions $\bm{\delta} \textbf{u}(t,\textbf{x})$ and $\bm{\delta \pi}(t,\textbf{x})$.
Likewise, the I\&BC $\{\textbf{v}_0; \textbf{v}_{\partial},\upeta_{\partial} \}$ can be written as:
\begin{equation}
\label{eq:localIBC}
 \textbf{v}_0(\textbf{x}) = \textbf{u}_0(\textbf{x}) + \bm{\delta} \textbf{u}_0(\textbf{x}),\quad 
  \textbf{v}_{\partial}(t,\textbf{s}) = \textbf{u}_{\partial}(t,\textbf{s}) + \bm{\delta} \textbf{u}_{\partial}(t,\textbf{s}), \quad
 {\upeta}_{\partial}(t,\textbf{s})= {\uppi_{\partial}}(t,\textbf{s}) + {\delta \uppi_{\partial}}(t,\textbf{s})
\end{equation}
with $\bm{\delta} \textbf{u}_0(\textbf{x})$, $\bm{\delta} \textbf{u}_{\partial}(t,\textbf{s})$ and ${\delta \uppi_{\partial}}(t,\textbf{s})$
sufficiently small so that $\{\textbf{v}_0;\textbf{v}_{\partial},\upeta_{\partial} \} \in \{ \textbf{u}_{0_{\langle \epsilon \rangle}}; 
\textbf{u}_{\partial_{ \ulcorner \epsilon \lrcorner}},\uppi_{\partial_{ \ulcorner \epsilon \lrcorner}}\}$.
Using Eq. \eqref{eq:localDesc}, the NSE for $\{\textbf{v},\bm{\eta}\}$ can be written as:
\begin{equation}
 \label{eq:localNSE}
 \textbf{u}_{,0} +\bm{\delta} \textbf{u}_{,0}  + (\textbf{u}\cdot \nabla)\textbf{u}+(\bm{\delta}\textbf{u}\cdot \nabla)\textbf{u} +(\textbf{u}\cdot \nabla)\bm{\delta}\textbf{u}+(\bm{\delta}\textbf{u}\cdot \nabla)\bm{\delta}\textbf{u}- \nu \nabla^2 \textbf{u}-\nu \nabla^2 \bm{\delta}\textbf{u} = -\bm{\pi} -\bm{\delta\pi} + \textbf{g}
\end{equation}
Since $\{\textbf{u},\bm{\pi}\}$ is already a solution of the NSE, the terms involving these fields cancel identically, resulting in
\begin{equation}
 \label{eq:localManifold}
 \bm{\delta} \textbf{u}_{,0}  +(\bm{\delta}\textbf{u}\cdot \nabla)\textbf{u} +(\textbf{u}\cdot \nabla)\bm{\delta}\textbf{u}+(\bm{\delta}\textbf{u}\cdot \nabla)\bm{\delta}\textbf{u}- \nu \nabla^2 \bm{\delta}\textbf{u} = -\bm{\delta\pi}
\end{equation}
with I\&BC $\{\bm{\delta} \textbf{u}_0;\bm{\delta} \textbf{u}_{\partial},{\delta \uppi_{\partial}}\}$.
In Eq. \eqref{eq:localManifold}, the function  $\textbf{u}(t,\textbf{x})$ should be considered a fixed point of the manifold $\mathfrak{F}_{\Upomega}$, not a dependent variable of the equation.
The solution $\{ \bm{\delta}\textbf{u},\bm{\delta}\bm{\pi}\}$ of Eq. \eqref{eq:localManifold} is such that $\{\textbf{u}+\bm{\delta}\textbf{u}, \bm{\pi}+\bm{\delta}\bm{\pi}\}$ is a solution of NSE \eqref{eq:NSE},
for the I\&BC \eqref{eq:localIBC}.
Eq. \eqref{eq:localManifold} serves to study the properties of the manifold $\mathfrak{F}_{\Upomega}$ in a neighbourhood of $\{\textbf{u},\bm{\pi}\}$, and thus its true dependent variable is $ \bm{\delta} \textbf{u}(t,\textbf{x})$.
Eq. \eqref{eq:localManifold} characterises the structure of such a neighbourhood and furnishes an additional insight into the properties of laminar and turbulent flows.

Assume now $\{\textbf{u},\bm{\pi}\}$ is a laminar flow.
According to Postulate \ref{pos:laminar}, $\bm{\delta} \textbf{u}(t,\textbf{x})$ would be small if $\epsilon$ is sufficiently small.
More precisely, for small $\delta\mathop{>}0$ there exists a small $\epsilon\mathop{>}0$ such that  $\bm{\delta} \textbf{u}\cong_{\delta} 0$.
In this case, the term $(\bm{\delta}\textbf{u}\cdot \nabla)\bm{\delta}\textbf{u}$
would be second order in $\delta$ and thus negligible.
In other words, if $\{\textbf{u},\bm{\pi}\}$ is a laminar flow, then the equation governing its neighbourhood would be
\begin{equation}
 \label{eq:localLinear}
 \bm{\delta} \textbf{u}_{,0}  +(\bm{\delta}\textbf{u}\cdot \nabla)\textbf{u} +(\textbf{u}\cdot \nabla)\bm{\delta}\textbf{u} - \nu \nabla^2 \bm{\delta}\textbf{u} = -\bm{\delta\pi}
\end{equation}
Note this equation is linear in $\{ \bm{\delta}\textbf{u},\bm{\delta}\bm{\pi}\}$,  i.e., if $\{ \bm{\delta}\textbf{u}_1,\bm{\delta}\bm{\pi}_1\}$ and $\{ \bm{\delta}\textbf{u}_2,\bm{\delta}\bm{\pi}_2\}$
satisfy Eq. \eqref{eq:localLinear}, then $\{ \lambda_1 \bm{\delta}\textbf{u}_1 + \lambda_2 \bm{\delta}\textbf{u}_2,\lambda_1\bm{\delta}\bm{\pi}_1\mathop{+}\lambda_2\bm{\delta}\bm{\pi}_2\}, \ \lambda_1,\lambda_2 \mathop{\in} \mathbb{R}$,
also satisfies it.
It follows that the neighbourhood of $\{\textbf{u},\bm{\pi}\}$ is diffeomorphic to a linear functional space, i.e., 
the functional space of solutions $\{ \bm{\delta}\textbf{u},\bm{\delta}\bm{\pi}\}$ of Eq. \eqref{eq:localLinear} constitute a Hilbert space
\footnote{Do not confuse the reported linear functional space in a neighbourhood of $\{\textbf{u},\bm{\pi}\}$
with the Euclidean tangent space existing at every point of any differentiable manifold.
The reported linear functional space exists only in the neighbourhood of a laminar flow, whereas the Euclidean tangent space exists everywhere in the manifold $\mathfrak{F}_{\Upomega}$.}.
The manifold $\mathfrak{F}_{\Upomega}$ of physical flows in $\Upomega$ is such that it contains a submanifold  $\mathfrak{F}_{\Upomega}^L$ of laminar flows,
which is locally diffeomorphic to a vector subspace of $H^2(\Upomega)$, and a submanifold  $\mathfrak{F}_{\Upomega}^T$ of turbulent flows with a local structure that is not a linear space.
The study of the local structure of $\mathfrak{F}_{\Upomega}$, which is an intrinsic geometrical property of the manifold, furnishes thus unambiguous criteria to assess the conditions under which flows are laminar or turbulent.
A parallel study may be conducted with the manifold of mean flows $\mathfrak{M}_{\Upomega}$, which brings forth identical results for the case of laminar flows.

\section{Turbulent steady-state flows}
\label{sct:steadyState}

A flow is called stationary or is in a steady state if the ensemble average of any physical field $\psi(t,\textbf{x})$ does not depend upon time 
\begin{equation}
\label{eq:steady}
\partial_0 \langle \psi(t,\textbf{x}) \rangle \equiv \Psi_{,0}(t,\textbf{x}) = \lim \limits_{N \rightarrow \infty} \frac{1}{N} \sum \limits_{n=1}^{N} \psi^{(n)}_{,0}(t,\textbf{x}) =0 \ \Rightarrow \ 	\Psi(\textbf{x})= \lim \limits_{N \rightarrow \infty} \frac{1}{N} \sum \limits_{n=1}^{N} \psi^{(n)}(t,\textbf{x})
\end{equation}
This definition is particularly remarkable when turbulent flows are involved.
It is not straightforward that a series of time-dependent terms yields a time-independent quantity.
The individual time-dependencies of each realisation must cancel one another during the averaging process, Eq. \eqref{eq:fluctAve}, i.e., 
the sum of the fluctuating components (time variations) of all realisations is a lesser infinite than $N$, which divides the sum (see Def. \ref{def:lesserInf} in \cref{sct:haps}).
A similar argument is applied if the mean field $\Psi(\textbf{x})$ is calculated through any of Eqs. \eqref{eq:ensembleAveNEQ}-\eqref{eq:ensembleAveChiralR}.

For any flow, stationary or not, it is possible to define a time average through the following limit
\begin{equation}
\label{eq:timeAver}
\Psi_T(\textbf{x}) :=  \lim \limits_{T \rightarrow \infty} \frac{1}{T} \int \limits_0^T  \psi(t,\textbf{x}) \dd t
\end{equation}
which is also a genuine average obeying the time-averaged RANSE (see \citep[Eq. (3.68)]{Gar17}).
This quantity is truly time-independent, by definition of the integral.
It is important to check whether $\Psi(\textbf{x})$ defined by Eq. \eqref{eq:steady} is equal to $\Psi_T(\textbf{x})$ defined by Eq. \eqref{eq:timeAver},
an equality that is usually taken for granted.

Let us consider a steady-state turbulent flow in the domain $\Upomega\mathop{=}[0,T_r] \mathop{\times} \Updelta \mathop{\subset} \mathbb{R}^4$,
physical I\&BC of range $\epsilon$, $\{ \psi_{0\langle \epsilon \rangle};\psi_{\partial \ulcorner \epsilon \lrcorner}  \}$,
and the ensemble of realisations $\{\psi^{(n)}(t,\textbf{x}) \}$ stemming from them, $\mathfrak{E}$.
This ensemble is used to calculate the mean field $\Psi(\textbf{x})$ of Eq. \eqref{eq:steady}.
Assume now that a similar steady-state turbulent flow is allowed to run for indefinite time $T\gg T_r$ in the open spatial domain $\Updelta$.
This flow could be coincident with a realisation of the ensemble during the interval $[0,T_r]$.
Divide the interval $[0,T]$ into disjoint consecutive intervals of identical duration $[t_{2n-1},t_{2n}), \ T_r =t_{2n}-t_{2n-1},\ n=1,2,3,...,N $,
such that the upper limit of the current interval is equal to the lower limit of next interval $t_{2n}=t_{2n+1}$.
Thus, we would have disjoint intervals $[t_1,t_2)=[0,T_r)$, $[t_3,t_4)$, $[t_5,t_6)$, $[t_7,t_8)$,... with $t_2=t_3$, $t_4=t_5$, $t_6=t_7$,...
The following mathematical development is possible:
\begin{align*}
& \Psi_T(\textbf{x}) =  \lim \limits_{T \rightarrow \infty} \frac{1}{T} \int \limits_0^T  \psi(t,\textbf{x}) \dd t =  	 \lim \limits_{N \rightarrow \infty} \frac{1}{N T_r} \sum \limits_{n=1}^{N}  \int \limits_{t_{2n-1}}^{t_{2n}}  \psi(t,\textbf{x}) \dd t \stackrel{(1)}{=}  \lim \limits_{N \rightarrow \infty} \frac{1}{N T_r}  \sum \limits_{n=1}^{N} \int \limits_{0}^{T_r}  \psi(t'+(n-1)T_r,\textbf{x}) \dd t'  \\
&\stackrel{(2)}{=}\lim \limits_{N \rightarrow \infty} \frac{1}{N T_r}   \sum \limits_{n=1}^{N} \int \limits_{0}^{T_r}  \psi_{(n)}(t',\textbf{x}) \dd t'
\stackrel{(3)}{=}\lim \limits_{N \rightarrow \infty} \frac{1}{ T_r}    \int \limits_{0}^{T_r} \frac{1}{N} \sum \limits_{n=1}^{N} \psi_{(n)}(t',\textbf{x}) \dd t'\stackrel{(4)}{=} \frac{1}{ T_r}  \int \limits_{0}^{T_r} \Psi(\textbf{x}) \dd t' =\Psi(\textbf{x})
\end{align*}
In (1) the general equality $t=t'+(n-1)T_r$, with $t \in [t_{2n-1},t_{2n})$ and $t'\in[0,T_r)$ is used, 
and a change of variable in the integral from $t$ to $t'$ is applied.
In (2) it is assumed that, for each $n\in \mathbb{N}$, $ \psi(t'+(n-1)T_r,\textbf{x})$ may be considered a sort of realisation of the flow within the interval $[0,T_r)$,
since $(n-1)T_r$ is constant for given $n$.
This pseudo-realisation is denoted by $\psi_{(n)}(t',\textbf{x})$ to distinguish it from a true realisation $\psi^{(n)}(t,\textbf{x})$.
In (3) the order of the sum and integral is interchanged, which is possible because $\psi$ is integrable and the series converges.
In (4) it is assumed that $\Psi(\textbf{x})$, as defined in Eq. \eqref{eq:steady}, is also equal to a similar series with $\psi_{(n)}(t',\textbf{x})$ instead of $\psi^{(n)}(t,\textbf{x})$.

The above derivation seems to demonstrate with elegance that $\Psi_T(\textbf{x})=\Psi(\textbf{x})$.
However, it might not be true, since in general $\psi_{(n)}(t',\textbf{x})$ would not be a realisation in the sense implied by the present research.
To begin with, each $\psi_{(n)}(t',\textbf{x})$ does not depart from an initial state within the physical initial conditions of range $\epsilon$, $\psi_{0\langle \epsilon \rangle}(\textbf{x})$,
but it is a slice of duration $T_r$ of a steady-state flow that has been evolving during a long time.
Furthermore, in general $\psi_{(n)}(t',\textbf{x})$ would not satisfy the requirements of Postulate \ref{pos:turbulent}, meaning that if $\epsilon$ is small enough, 
then it cannot be found any time $\tau>0$ such that $\psi_{(n)} \stackrel{_\tau}{\cong}_{\epsilon} \psi_{(m)}$, for any two pseudo-realisations $n$ and $m$, see Eq. \eqref{eq:almostEqualT}.
Otherwise put, we may have 
$\psi_{(n)}(t',\textbf{x})\equiv \psi(t'+(n-1) T_r,\textbf{x}) \not \approx  \psi(t'+(m-1) T_r,\textbf{x}) \equiv \psi_{(m)}(t',\textbf{x}), \ 0\leq t'<\tau$, and in particular 
$\psi_{(n)}(0,\textbf{x}) \not \approx \psi_{(m)}(0,\textbf{x})$,
which is contrary to what is required by the initial conditions.
Therefore, if $\psi_{(n)}(t',\textbf{x})$ does not generally qualify as a realisation in the sense implied by this research,
then it might well occur that step (4) be false, since $\Psi(\textbf{x})$ would not be equal to the series of $\psi_{(n)}(t',\textbf{x})$.
The conclusion is that, in general, $\Psi_T(\textbf{x}) \neq \Psi(\textbf{x})$, which would perhaps come as a surprise to many readers, albeit it might well be that their difference be negligible.

It may be argued that the situation just described is unavoidable in turbulent flows, for it is not obvious how could one prepare two realisations that begin with almost identical conditions,
as it is demanded in the statement ``\textit{turbulent realisations departing from physical initial conditions of range $\epsilon$}''.
Certainly, accomplishing such a task is not particularly difficult in the realm of computer simulations.
In the realm of experimental fluid mechanics, though, possibly the most reliable method to generate suitable realisations would be to begin all experiments in laminar regime.
In any case, the present work is a theoretical research and cannot say much about experimental practice.

\section{Finite number of realisations}
\label{sct:finiteRealis}

The rigorous properties of ensemble-averaged fields stem from the fact that $N \rightarrow \infty$, and $N$ is dividing the sum defining the average,
and thus any feature of realisations not sufficiently repeatable will be ironed out after division by $N$.
However, in practical terms, the condition $N \rightarrow \infty$ is unattainable and most often $N$ would be reported as less than one hundred.
This section will propose an approximate method to determine the minimum value of $N$ that would still ensure sufficiently accurate mean fields.

First, the ensemble average of the field $\psi(t,\textbf{x})$ for a finite number $N$ of realisations, also called finite-ensemble average or finite mean, is defined as:
\begin{equation}
 \label{eq:finiteNMean}
 \Psi_{(N)}(t,\textbf{x}):=\frac{1}{N}
\sum \limits_{n=1}^{N}  \psi^{(n)}(t,\textbf{x}) \ , \qquad (t,\textbf{x}) \in \Upomega 
\end{equation}
and akin expressions if the finite-mean field is calculated through any of Eqs. \eqref{eq:ensembleAveNEQ}-\eqref{eq:ensembleAveChiralR}.
In particular, $\Psi_{(\infty)}(t,\textbf{x}) \equiv \Psi(t,\textbf{x})$.
The announced method to estimate the minimum number $N$ of realisations is based on Cauchy's convergence criterion (see \citep{BCP21}), and it unfolds as follows:
(i) Select the physical I\&BC of range $\epsilon$, $\{\psi_{0\langle \epsilon \rangle}; \psi_{\partial \ulcorner \epsilon \lrcorner}\}$, 
from which the ensemble $\mathfrak{E}$ will be generated.
(ii) Consider a sufficiently small maximum admissible error $\epsilon_0 >0$.
(iii) Execute $N-3$ realisations of the same flow, according to the details offered in \cref{sct:ensembAve}, and calculate $\Psi_{(N-3)}(t,\textbf{x})$ from Eq. \eqref{eq:finiteNMean}.
(iv) Execute the realisation number $N-2$ and calculate  $\Psi_{(N-2)}(t,\textbf{x})$.
(v) Check if the following condition is satisfied: 
\begin{equation}
 \label{eq:quotientN-2}
 \frac{\norm{\Psi_{(N-2)}-\Psi_{(N-3)}}}{\norm{\Psi_{(N-3)}}} < \epsilon_0
\end{equation}
whereby the norm $\norm{\cdot}$ is defined by Eq. \eqref{eq:norm}.
Let us assume that condition \eqref{eq:quotientN-2} is satisfied.
(vi) Repeat steps iv and v: Execute the realisation number $N-1$, calculate $\Psi_{(N-1)}(t,\textbf{x})$ and check if 
\begin{equation}
 \label{eq:quotientN-1}
 \frac{\norm{\Psi_{(N-1)}-\Psi_{(N-2)}}}{\norm{\Psi_{(N-2)}}} < \epsilon_0
\end{equation}
still holds.
Let us assume, again, that condition \eqref{eq:quotientN-1} is met.
(vii) Repeat, one last time, steps iv and v: Execute the realisation number $N$, calculate $\Psi_{(N)}(t,\textbf{x})$ and check if 
\begin{equation}
 \label{eq:quotientN}
 \frac{\norm{\Psi_{(N)}-\Psi_{(N-1)}}}{\norm{\Psi_{(N-1)}}} < \epsilon_0
\end{equation}
is still satisfied.
If condition \eqref{eq:quotientN} is also fulfilled, then $N$ should be a sufficient size for the ensemble of realisations.
Otherwise, repeat the process until three consecutive finite means satisfy conditions \eqref{eq:quotientN-2}-\eqref{eq:quotientN}.

The rationale supporting this method is the following: If three consecutive additional realisations, divided respectively by $N-2$, $N-1$ and $N$,
contribute very little to the finite mean (less than $\epsilon_0$), then it is very likely that the finite mean has already entered into asymptotic zone,
and further realisations would contribute even less to the finite mean, since with each new realisation the denominator increases.
The particularities of individual realisations, after being divided by increasing $N$, become decreasingly influential in the finite mean obtained through this procedure.
A study about how such particularities affect the mean fields is included in \cref{sct:haps}.

\section{Conclusions}
\label{sct:Conclusions}

An unambiguous mathematical definition of laminar and turbulent flows has been proposed.
It is based on an experimental truth: A laminar flow can be repeated indefinitely with essentially the same results (chiral flows notwithstanding),
whereas each repetition of a turbulent flow will always give different results.
This basic fact leads to mean fields that are quasi-solutions of the NSE \eqref{eq:NSE} if the flow is laminar,
or to mean fields that are nowhere near a solution of the NSE when the flow is turbulent.
Otherwise put, the mean turbulent flow is not physical, it cannot naturally occur in any experiment.
Laminar flows arise in situations of restricted nonlinearity, Eq. \eqref{eq:condLaminarEq}, whereas turbulent flows are the consequence of general nonlinearity.
This is essentially the mathematical definition proposed in the present research, which can be used to conceive and prove theorems concerning laminar and turbulent flows. 

Being the solution (or quasi-solution) of a given equation is a clear and unambiguous criterion to define a dynamical system, which is usually employed in physics.
And conversely, not being a solution also serves to define other types of systems.
Therefore, the proposed definition follows the long-standing method so successfully applied in physics.

In summary, the following procedure to know unambiguously whether a particular flow is laminar or turbulent is proposed: 
(i) give up scrutinising the individual characteristics of such a particular flow,
(ii) fix the degree of approximation $\epsilon$ and define a set of $\epsilon$-equal I\&BC,
(iii) run as many realisations as possible, all starting $\epsilon$-identically and being subject to $\epsilon$-identical conditions (the ensemble), and let each realisation follow its own dynamics,
(iv) calculate the mean fields according to Eqs. \eqref{eq:ensembleAve}-\eqref{eq:ensembleAveChiralR}, as required by the type of flow,
(v) if the mean fields are almost identical to those of the realisations, then the particular flow is laminar; otherwise it is turbulent.
Alternatively, (v) if the mean fields are a quasi-solution of the NSE, then the particular flow is laminar; otherwise it is turbulent.

Furthermore, for $\epsilon$ sufficiently small, the set of laminar realisations contained in the ensemble $\mathfrak{E}_{\epsilon}$ constitutes a class of equivalence,
i.e., to all practical effects, all those physical flows represent the same laminar flow, since they are indistinguishable if examined in a laboratory.
Even the mean laminar flow would be a representative of the equivalence class, since it is arbitrarily proximate to the realisations.
Conversely, the realisations of a turbulent flow in an ensemble $\mathfrak{E}_{\epsilon}$ cannot constitute an equivalence class, will not be indistinguishable from one another, regardless of $\epsilon >0$.

Possibly, some readers may think that these ideas were already part of the pool of conventional wisdom existing in fluid mechanics.
Be that as it may, to the best of our knowledge, an approach such as that offered herein cannot be found in the existing literature, 
it has not been disclosed before and, therefore, deserves to be exposed to the community.
Arguably, after reading this work, some researchers will eventually venture other definitions that might even refine and improve ours,
because there are other properties that distinguish laminar from turbulent flows.
Hopefully, the new path revealed here will serve as an inspiration for further fruitful research.
Once the floodgates have been opened, there is no stopping the tide.

\section*{Declaration of interests}
The authors report no conflict of interest.

\section*{Author contribution}
\label{sct:Contributions}

FJGG: Conceptualization, investigation, methodology, mathematical development, data gathering, figures, manuscript writing (initial draft) and editing.

PFA: Supervision, critical revision, validation, editing and funding acquisition.

\bibliographystyle{alpha}
\bibliography{J118_02R00_Formal_Definition_Turbul_Bibliography}  %

\section*{Appendices}

\appendix

\section{A simple non-mathematical introduction}
\label{sct:simpleIntro}

As stated in the title, the present research aims to provide a non-ambiguous general mathematical definition of laminar and turbulent flow,
which could be used to prove theorems, generate new definitions and construct new theories.
Some readers may find it difficult to ultimately understand what is proposed in this paper, since the mathematical language often turns very abstract.
For those readers, we shall try to summarise the basic ideas that constitute the essence of the definition.

For simplicity, we shall assume that a flow becomes fully characterised by its velocity field alone, ignoring other quantities such as the pressure.
Only incompressible Newtonian flows are considered in this research.

Flows come in two flavours.
On one hand, we have the physical flow $\textbf{u}(t,\textbf{x})$ that we can touch and examine in laboratory, industry or natural environment.
On the other hand, we have the average or mean flow $\textbf{U}(t,\textbf{x})$, which is devoid of the minutiae and particularities of the individual physical flows 
(we should not be concerned now about how the average flow is obtained).
The mean flow is not physical, it is a construction of the mind that requires a mathematical procedure to emerge.

Given any flow, we can always presume the existence of its fields $\textbf{u}(t,\textbf{x})$ and $\textbf{U}(t,\textbf{x})$, 
the actual physical flow and the mean flow that would be derived from it.
The physical flow $\textbf{u}(t,\textbf{x})$ satisfies the NSE \eqref{eq:NSE},
while the mean flow $\textbf{U}(t,\textbf{x})$ satisfies the RANSE \eqref{eq:RANSE}.
The mathematical correspondence is simple: $\textbf{u}(t,\textbf{x})$ with the NSE and $\textbf{U}(t,\textbf{x})$ with the RANSE, period.

The definition of laminar and turbulent flow focuses exclusively on $\textbf{U}(t,\textbf{x})$, and is surprisingly simple:
If $\textbf{U}(t,\textbf{x})$ happens to be also a solution of the NSE (actually, a quasi-solution), then the flow is laminar.
However, if $\textbf{U}(t,\textbf{x})$ is not a quasi-solution of the NSE, then the flow is turbulent.
In either case, $\textbf{U}(t,\textbf{x})$ is always a solution of the RANSE.
Recall that being a quasi-solution means to be arbitrarily proximate to an actual solution.

Note that the notion of being solution of a differential equation is mathematically unambiguous and perfectly correct.
There is nothing to object to the definition of laminar flow from the mathematical standpoint.
Likewise, the notion of not being solution of a differential equation is also not ambiguous and sensible.
Therefore, there is nothing to reproach in the proposed mathematical definitions.

The remainder of the paper is devoted to describing and justifying the criteria and procedures for generating an ensemble, calculating mean fields from the ensemble, and exposing various results (theorems) that are equivalent to the proposed definitions.
The reader should consider them as necessary intermediate steps to reach the rigorous mathematical definitions included in the paper.
Hopefully, the reader will now be in a better position to evaluate our research.
The next two solved examples will further illustrate the capabilities and methods of our theory; the reader will see that it is not so difficult to apply it to practical problems.

\section{A first example: The von K\`arm\`an vortex street}
\label{sct:KarmanVortex}

It must be proved, following the definitions proposed herein, that the flow commonly known as the von K\`arm\`an vortex street is laminar.
\begin{figure}[h]
	\begin{center}
		\leavevmode
		\includegraphics[width=0.85 \textwidth, trim = 0mm 0mm 0mm 0mm, clip=true]{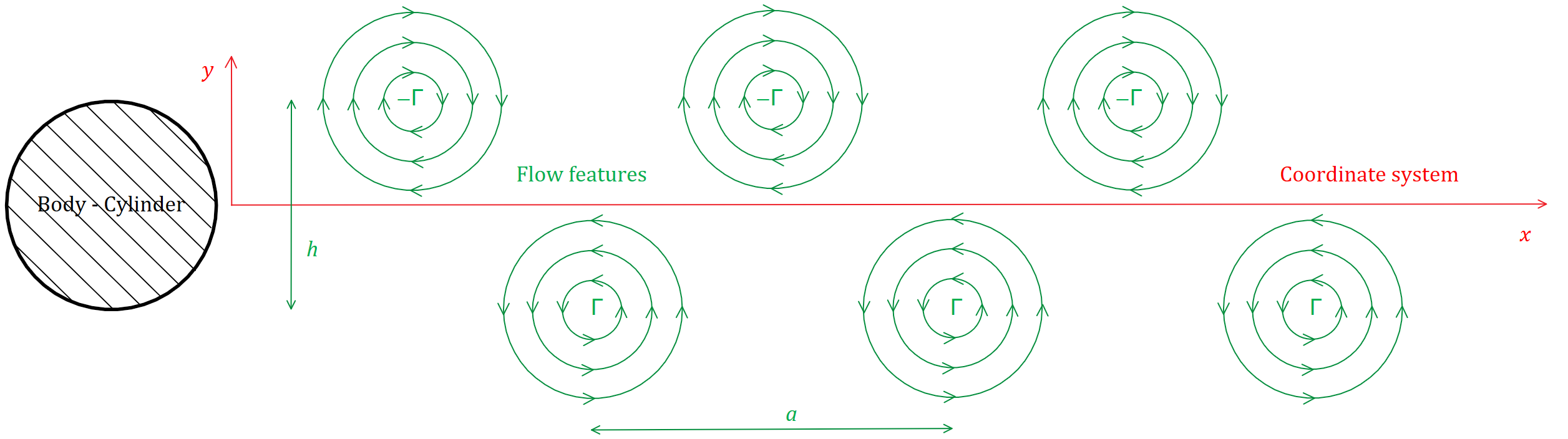}
		\caption{\small{Sketch of von K\`arm\`an vortex street.}}
		\label{fig:KarmanStreet}
	\end{center}
\end{figure}
To proceed with the demonstration, we formally need a huge collection of realisations, from which to gather the corresponding ensemble and all mean fields derived from it. 
Another possibility would be departing from the exact analytical solution of the problem, which, to our knowledge, is not yet available. 
Therefore, we have to make do with an approximate analytical solution and \cite{AK62} offers a very good one.
No attempt will be made to explain the analytical solution; the interested reader is addressed to the reference \cite{AK62}, whose nomenclature is used throughout.

It should be noted that viscosity only plays a significant role in the generation of the vortex layers around the blunt body that causes the vortex shedding,
and it can be neglected in the formation of the vortex street created from the vortex layers.
Thus, the inviscid hypothesis assumed by \cite{AK62} is quite close to reality, provided it is not applied to the proximity of the cylinder causing the vortex street.
The suitability of the model offered in \cite{AK62} is supported by its Figs. 5 to 14, which show a convincing evolution of the von K\`arm\`an vortex street.

We shall assume the model proposed in \cite{AK62} without discussion.
Suffices to say that \cite{AK62} supposes two parallel vortex sheets being generated by the cylinder under a freestream of velocity $U$,
that the vortex sheets undergo a small sinusoidal perturbation, which decompose them into $n$ point vortices per sheet and per wavelength.
These point vortices will then give rise to the vortex street as depicted in Fig \ref{fig:KarmanStreet}.
Since our purpose is to confirm that the vortex street is laminar, we shall use the mathematical functions reported in \cite{AK62} as equivalent to a realisation of the flow.

The flow is assumed two-dimensional, inviscid and incompressible.
It follows that all vorticity vectors are perpendicular to the paper.
More importantly, the vorticities can be added as scalars, since all have the same direction.
Conventionally, the arrangement of Fig. \ref{fig:KarmanStreet} is called ``L-configuration'',
with the first vortex in the upper row.
A symmetric arrangement with the first vortex in the lower row is called ``R-configuration''.
The two-dimensional velocity $(u_i,v_i)$ of the $i^{th}$ upper vortex in any realisation is given by \cite{AK62}
\begin{equation}
 \label{eq:velCompX}
 u_i=U+\frac{U}{n} \sum \limits_{j=1}^n \left( \frac{\sinh 2\pi(y_i-y_j)/a}{\cosh 2\pi (y_i-y_j)/a-\cos 2\pi(x_i-x_j)/a } - \frac{\sinh 2\pi(y_i+y_j)/a}{\cosh 2\pi (y_i+y_j)/a +\cos 2\pi(x_i-x_j)/a }  \right)
\end{equation}
\begin{equation}
 \label{eq:velCompY}
 v_i=-\frac{U}{n} \sum \limits_{j=1}^n \left( \frac{\sin 2\pi(x_i-x_j)/a}{\cosh 2\pi (y_i-y_j)/a-\cos 2\pi(x_i-x_j)/a } + \frac{\sin 2\pi(x_i-x_j)/a}{\cosh 2\pi (y_i+y_j)/a +\cos 2\pi(x_i-x_j)/a }  \right)
\end{equation}
The functions above are analytic, since $\cosh x > 1$ for all $x\neq 0$ and the denominators are either never zero, or zero only if $i=j$, in which case numerators are also zero, leaving quotients finite.
From Theorem \ref{thr:analyticity} follows that the flow modelled by Eqs. \eqref{eq:velCompX}-\eqref{eq:velCompY} is laminar, and this would end the demonstration.
However, it is an interesting exercise to demonstrate explicitly that this flow is laminar, so that readers may learn the details of the proposed method.

We proceed now to execute many realisations of this flow, each with very small variations respect to that shown in Fig. \ref{fig:KarmanStreet}.
Most likely, we shall get realisations of both chiralities, $L$ and $R$, which should be ensembled separately.
Now, suppose two realisations $k$ and $m$ of the same chirality.
For each one, we would obtain freestream velocities $U^{(k)}$ and  $U^{(m)}$, geometrical parameters $a^{(k)}$ and $a^{(m)}$,
and velocities $(u_i^{(k)},v_i^{(k)})$ and $(u_i^{(m)},v_i^{(m)})$, all of them very close to each other.
We shall perform the calculations with only these two realisations; the case of $N$ realisations is a straightforward generalisation. 
Suffices to take one of the terms of Eqs. \eqref{eq:velCompX}-\eqref{eq:velCompY} and calculate its ensemble average; for other terms the procedure is similar.
The mean-velocity field can be calculated using the Taylor series method outlined in Theorem \ref{thr:analyticity}.
However, in this case we shall employ a somewhat different method, equally effective.

For any continuous function $f(x)$ the intermediate value theorem holds: Let $x_1<x_2$ be two points in the domain of $f(x)$, then for every $y$ such that $\min( f(x_1),f(x_2))<y<\max(f(x_1),f(x_2))$ there exists $\bar{x}\in[x_1,x_2]$ such that $y=f(\bar{x})$.
This holds particularly for the mean value $y=(f(x_1)+f(x_2))/2$.
This same principle is valid, approximately, for any of the terms in Eqs. \eqref{eq:velCompX}-\eqref{eq:velCompY} if it is applied to the parameters $\{U,a\}$, instead of the variables $(x_i,y_i)$.
Note all terms depend continuously on the parameters $\{U,a\}$ and $a>0$.
If $U^{(k)} \approx U^{(m)}$ and  $a^{(k)} \approx a^{(m)}$ are sufficiently proximate to each other,
then it is possible to find $\bar{U}\in[\min( U^{(k)} ,U^{(m)}) ,\max( U^{(k)} ,U^{(m)})] $ and $\bar{a}\in[\min(a^{(k)} ,a^{(m)}), \max(a^{(k)} ,a^{(m)}) ]$
such that the average satisfies:
\begin{align}
 \label{eq:average}
 \frac{1}{2} & \left( \frac{U^{(k)} \sinh 2\pi(y_i-y_j)/a^{(k)}}{\cosh 2\pi (y_i-y_j)/a^{(k)}-\cos 2\pi(x_i-x_j)/a^{(k)} }  +
 \frac{U^{(m)} \sinh 2\pi(y_i-y_j)/a^{(m)}}{\cosh 2\pi (y_i-y_j)/a^{(m)}-\cos 2\pi(x_i-x_j)/a^{(m)} }  \right) \approx \notag \\
 & \quad  \frac{\bar{U} \sinh 2\pi(y_i-y_j)/\bar{a}}{\cosh 2\pi (y_i-y_j)/ \bar{a}-\cos 2\pi(x_i-x_j)/\bar{a} }  
\end{align}
for the first term within the sum of Eq. \eqref{eq:velCompX} and likewise for remaining terms.
Here, the important point is that averaged terms are arbitrarily proximate to functions that have exactly the same functional form as the realisation's terms.
Since $U^{(k)} \approx U^{(m)}$ and  $a^{(k)} \approx a^{(m)}$, then $\bar{U} \approx U^{(k)} \approx U^{(m)}$ and  $\bar{a} \approx a^{(k)} \approx a^{(m)}$,
which entails the condition of laminar flow, see Def. \ref{def:laminarFlow}.
A similar calculation should be executed for the remaining terms, including the $i^{th}$ lower vortices.

This procedure must be repeated for $N$ realisations, not just two, and the end result will be similar: 
The mean values $\langle u_i \rangle$ and $\langle v_i \rangle$ are arbitrarily proximate to the realisation characterised by $\bar{U}, \bar{a}$.
Therefore, if the mean flow has (almost) the same analytical form as any realisation, it follows that the mean flow should be a quasi-solution of the NSE,
because any realisation (physical flow) is already a solution of the NSE.
The conclusion offers no doubt: The von K\`arm\`an vortex street is a laminar flow, according to the definition proposed herein.

\section{A second example: Laminar flow with imposed white noise}
\label{sct:whiteNoise}

Now follows a very interesting example that shows efficiently how the proposed definitions permit to classify the flows into their correct classes, laminar or turbulent.
We shall purposely restrict ourselves to a simple case, just enough to illustrate our point.
Similar, more complex situations can also be posed and readers are invited to think of their own.

Imagine a duct, for example an arbitrarily long pipe of radius $R$, in which a low Reynolds number steady flow is developed with axial velocity profile $u_s(r,z)$, 
where $r$ is the radial coordinate and $z$ the axial coordinate in a cylindrical system $(r,\theta,z)$ 
(we assume circular symmetry, i.e. the velocity does not depend on the azimuth $\theta$).
The flow is created by a constant pressure gradient $\Pi$, and a constant velocity $U_0$ is assumed everywhere at the inlet section ($z=0$).
What is observed is that the initially homogeneous velocity $U_0$ changes as the fluid advances along the pipe, and several metres downstream the flow becomes fully developed,
i.e., the velocity profile acquires the parabolic form of a Hagen-Poiseuille flow
\begin{equation}
 \label{eq:HP} 
 u_s(r) = -\frac{\Pi}{4 \mu} \left(1- \frac{r^2}{R^2}  \right) 
\end{equation}
Now imagine that at the inlet section of the pipe we apply an additional velocity perturbation $\textbf{u}'_0(t,r,\theta,0)$ with the spectrum of white noise,
i.e., the velocity field at the inlet is 
\begin{equation}
 \label{eq:whiteNoise}
 \textbf{u}_0(t,r,\theta,0)=(0,0,U_0)+\textbf{u}'_0(t,r,\theta,0)
\end{equation}
The perturbation could be extensive, covering the whole inlet section, or it could be of limited extension, such as that caused by a loudspeaker.
Either way, a fluctuating velocity field $\textbf{u}(t,\textbf{x})$ is produced in the initial portion of the pipe, called the entrance region.
Since the Reynolds number is low, what it is observed is that the perturbation will gradually dissipate along the pipe and several metres downstream the flow will be laminar,
with the parabolic velocity profile typical of Hagen-Poiseuille flow, Eq. \eqref{eq:HP}.
The mean flow is steady, that is, the statistical moments derived from the flow quantities are time independent.
The white-noise perturbation is supposed to have a fairly constant non-zero r.m.s. value, and its duration is very long. 

\

The described flow is laminar or turbulent?
Let us apply to it both, the conventional notion of turbulence and the new definitions proposed in this work, to learn what each one has to say.

\textbf{CONVENTIONAL NOTION OF TURBULENT FLOW}

The described flow is indeed fluctuating and time dependent.
The procedure associated to the conventional notion of turbulence requires that we first calculate a time-mean velocity field (or even a spatial-mean)
\begin{equation}
 \label{eq:timeMean}
 \bar{\textbf{u}}(\textbf{x}) = \lim \limits_{T\rightarrow \infty} \frac{1}{T} \int \limits_0^T \textbf{u}(t,\textbf{x}) \dd t
 \end{equation}
such that the instantaneous velocity field $\textbf{u}(t,\textbf{x}) =\bar{\textbf{u}}(\textbf{x})+ \textbf{u}'(t,\textbf{x})$ 
be expressed in terms of a Reynolds decomposition.
Note $\bar{\textbf{u}}(\textbf{x})$ cannot be zero (except at the wall), because the fluid is moving downstream on average. 
The fluctuating component $\textbf{u}'(t,\textbf{x})$ has a non-zero r.m.s. value, and thus the turbulence intensity exists and is positive:
\begin{equation}
 \label{eq:turbIntensity}
 I = \frac{u'_{rms}}{|\bar{\textbf{u} }|} > 0
\end{equation}
The turbulence intensity is not negligible, at least in the upstream portion of the pipe that is called the entrance region. 
Therefore, the conventional notion of turbulence dictates that the described flow be turbulent, despite having a low Reynolds number.

\textbf{PROPOSED MATHEMATICAL DEFINITION OF LAMINAR AND TURBULENT FLOW}

Let us see what the newly proposed definition has to say.
A typical realisation of this flow would be $\textbf{u}^{(n)} (t,\textbf{x})$.
Since it is a realisation, i.e., a true physical flow, it must be a solution of the NSE with the I\&BC described in the problem.
In particular, the boundary condition at the inlet is the sum of a constant axial velocity $(0,0,U_0)$ plus a white-noise perturbation $\textbf{u}'^{(n)}_0(t,r,\theta,0)$, see Eq. \eqref{eq:whiteNoise}.
Let us run another realisation $\textbf{u}^{(m)} (t,\textbf{x})$ with a similar boundary condition at the inlet $(0,0,U_0)+\textbf{u}'^{(m)}_0(t,r,\theta,0)$.
Since the boundary condition is white noise, we can safely assume that $\textbf{u}'^{(n)}_0(t,r,\theta,0)$ and $\textbf{u}'^{(m)}_0(t,r,\theta,0)$ 
are boundary quasi-equal of range $\epsilon$, that is, the integral of Eq. \eqref{eq:almostBoundEqual0} is fulfilled for some small $\epsilon$, 
with $f=\textbf{u}'^{(n)}_0$ and $g=\textbf{u}'^{(m)}_0$. 
We have built our ensemble.

We calculate now the mean fields that are inferred from the ensemble, beginning with the boundary conditions.
The white-noise part of the mean boundary condition at the inlet is
\begin{equation}
 \label{eq:BCWhiteNoise}
 \textbf{U}'_0(t,r,\theta,0) = \lim \limits_{N\rightarrow \infty} \frac{1}{N} \sum \limits_{n=1}^N \textbf{u}'^{(n)}_0(t,r,\theta,0)
\end{equation}
It is well known that the mean value of white noise is zero (not the r.m.s.); therefore, $\textbf{U}'_0(t,r,\theta,0)=0$.
It follows from Eq. \eqref{eq:whiteNoise} that the mean boundary condition at the inlet is reduced to $\textbf{U}_0(t,r,\theta,0)=(0,0,U_0)$.
But this is exactly the flow $u_s(r,z)$ described at the beginning, before the white-noise perturbation was applied.
The mean flow $\textbf{U}(t,\textbf{x})$ actually becomes $(0,0,U_s(r,z))$ and is formally identical to a laminar realisation that begins with a constant axial velocity $U_0$ at the inlet,
and ends with a parabolic Hagen-Poiseuille profile $U_s(r)$ identical to Eq. \eqref{eq:HP}.
Our definition undoubtedly dictates that the flow must be classified as laminar.

\

The contradiction between the conventional approach and our theory is only apparent, because the described flow is indeed laminar, it has been laminar from start to finish.
The flow appears to be turbulent because it is fluctuating, but it is so because the boundary conditions are fluctuating.
Note the white-noise perturbation is a boundary condition; it cannot be otherwise, because we cannot prescribe a velocity field when solving the NSE, we can only prescribe the I\&BC.
The flow is simply following the boundary condition; it did the best it could to cope with the imposed fluctuation.

The flow can not ignore the imposed fluctuation and has to fluctuate with it, although it managed to dissipate the fluctuation downstream.
The own dynamics of the flow tends to damp the prescribed fluctuation, and in the end the flow becomes a Hagen-Poiseuille further downstream.
This damping of the imposed fluctuation is the best symptom that the flow is laminar, despite the fluctuating behaviour that is inherited, and which is not self-generated by the dynamics.

For example, if we have a large deposit high enough to create a great constant pressure gradient, the flow will be fluctuating (turbulent) despite the pressure gradient being strictly constant.
The dynamics of the flow forces the velocity to fluctuate, despite the source of motion (the pressure gradient) being constant.
This is a typical behaviour of turbulent flow.
However, this is not the case in our example.
The velocity does not fluctuate because its dynamics dictates so, it fluctuates because it has a prescribed white-noise boundary condition.
But, as soon as the flow can, the fluctuation dissipates and it ends up being non-fluctuating, i.e., apparently laminar.
The truth is that the flow has not ceased to be laminar.
The flow has dealt with the prescribed fluctuation as best it can, first by fluctuating itself, then by gradually reducing the fluctuation until, 
further downstream, it ceases to fluctuate at all.

Our definition is not fooled by the apparent turbulence indicated by a highly variable fluctuating component.

\section{A brief theory of haps}
\label{sct:haps}

This section is devoted to study how and when the particular features of realisations are transmitted to the corresponding mean fields of the ensemble.
The mean fields that emerge from the ensemble of realisations would not be properly characterised without considering the influence of individual attributes of physical fields upon them.
We are going to study the conditions under which potential particularities of realisations affect the ensemble-averaged quantities and fields.
To this end, the notion of realisation hap, or simply a hap, must be introduced.

A hap $\mathfrak{H}$ is some identifiable turbulent phenomenon or structure that occurs in a realisation. 
It could be an eddy, a coherent structure, a blob, a puff, a streak, a roll, a slug... or any other of those imaginative names endowed to the nonlinear phenomena associated with turbulence.
Examples of haps would be the occasional turbulent spots occurring in the transition from laminar to turbulent flow, the wake behind a solid body immersed in a fully turbulent flow,
the separation region in a backward-facing step flow, the thermal plume shed by a hot body, etc.
Haps always occur in physical flows and can be measured, i.e., they are natural events related with turbulence.
A hap $\mathfrak{H}$  occurring during the $n^{th}$ realisation is characterised by the following intrinsic properties: 
(i) a temporal extension given by the time interval $B^{(n)} \mathop{=} (t_0^{(n)},t_0^{(n)}+\delta t^{(n)} )\subset [0,T_r]$,
(ii) a time-dependent extension in space $\delta \Updelta^{(n)}_t$, $t \in B^{(n)}$,
(iii) an evolutionary spacetime domain $\mathfrak{H}_{\upomega}^{(n)} \subset \Upomega$, which should be understood as a fibre bundle with base $B^{(n)}$ 
and fibre $ \delta \Updelta^{(n)}_t$ for each $t \in B^{(n)}$ (see \cite{Ste51}),
(iv) a geometric centroid $\textbf{x}_{\mathfrak{H}}^{(n)}(t)$, $t \in B^{(n)}$, i.e., the hap's time-dependent centre of mass, 
and (v) a set of physical fields $\psi^{(n)}(t,\textbf{x})$ whose variations can be measured to determine whether the hap is occurring 
(typically, $\psi^{(n)}(t,\textbf{x})$ would be the velocity, pressure or vorticity fields).

The hap is characterised by the following condition: Within the hap's spacetime domain $\mathfrak{H}_{\upomega}^{(n)} \subset \Upomega$ of any realisation $n$, the physical field $\psi^{(n)}(t,\textbf{x})$ may be expressed as 
\begin{equation}
 \label{eq:hapDecomp}
 \psi^{(n)}(t,\textbf{x}) =  \psi^{(n)-}(t,\textbf{x}) +  \psi^{(n)}_h(t,\textbf{x})\ , \quad \forall \ (t,\textbf{x}) \in \mathfrak{H}_{\upomega}^{(n)} 
\end{equation}
where $ \psi^{(n)-}(t,\textbf{x})$ is called the hapless field and $ \psi^{(n)}_h(t,\textbf{x})$ the hap field.
The decomposition \eqref{eq:hapDecomp} is not like the Reynolds decomposition of Eq. \eqref{eq:ReynoldsDecomp}.
The hapless field would be the value taken by $\psi(t,\textbf{x})$ if the hap were non-existent, whereas the hap field is what permit us to affirm the hap exists.
Inside the hap, the hapless field $\psi^{(n)-}(t,\textbf{x})$ is like a background over which the hap itself develops, it is akin to the existing physical field outside the hap,
and the hap field $\psi^{(n)}_h(t,\textbf{x})$ makes the difference between the hap and its surroundings.
For example, in a Hagen-Poiseuille laminar flow which is on the verge of transition to turbulence, the hap would be a turbulent puff (or whatever the name) occurring in the $n^{th}$ realisation, 
with a measurable time duration and spatial extent.
The physical field would be the velocity, which within the hap would be time-dependent and three-dimensional $\textbf{u}^{(n)}(t,\textbf{x})$, while outside the hap would be steady and unidirectional, 
parallel to the pipe's axis, $(0,0,u^{(n)}(r))$ in cylindrical coordinates $(r,\varphi,z)$, 
with $u^{(n)}(r)$ equal to the Hagen-Poiseuille parabola.
The hapless field would be the value taken by the velocity if the hap were non existent, i.e., $\textbf{u}^{(n)-}(t,\textbf{x})=(0,0,u^{(n)}(r)), \  (t,\textbf{x}) \in \mathfrak{H}_{\upomega}^{(n)}$, which is time independent.
The hap field would be the difference, i.e., what makes the hap exists, which is the time-dependent three-dimensional field $\textbf{u}^{(n)}_h(t,\textbf{x})=\textbf{u}^{(n)}(t,\textbf{x})-(0,0,u^{(n)}(r)), \  (t,\textbf{x}) \in \mathfrak{H}_{\upomega}^{(n)}$.
Note this decomposition only makes sense inside the hap; outside the hap the hapless field coincides with the physical field, 
and it is sufficient in itself to characterise the flow and need not be decomposed, 
$\psi^{(n)}(t,\textbf{x}) \mathop{=}  \psi^{(n)-}(t,\textbf{x})$, $ \psi^{(n)}_h(t,\textbf{x}) =0$, $\forall \ (t,\textbf{x}) \notin \mathfrak{H}_{\upomega}^{(n)}$.

In general, regrettably, the role of hapless and hap fields will not be so straightforward as in this example.
The decomposition of Eq. \eqref{eq:hapDecomp} may appear to the reader as not sufficiently rigorous. 
In principle, there is not a unique way to accomplish the decomposition $\psi^{(n)}=\psi^{(n)-}+\psi^{(n)}_h$, 
since it might not be unambiguous where or when the hap begins or ends, or which would be a correct hap field.
However, we shall see that all results reported in this section hold for any physically meaningful decomposition of the field $\psi$.
The theory purports to explain how the characteristics of individual haps are transmitted to the ensemble-averaged fields, regardless of the particular decomposition considered (within reason).

A hap $\mathfrak{H}$ may not occur in every realisation. 
We shall say that a hap $\mathfrak{H}$ is not occurring in the $n^{th}$ realisation if $\mathfrak{H}_{\upomega}^{(n)} $ is a null-measure domain ($\upmu(\mathfrak{H}_{\upomega}^{(n)})=0$).
Note that the non-occurrence of a hap $\mathfrak{H}$ in the $n^{th}$ realisation does not imply that  $\mathfrak{H}_{\upomega}^{(n)}$ be empty ($\mathfrak{H}_{\upomega}^{(n)}=\emptyset$);
it simply means that it has no measurable duration ($\upmu_1(B^{(n)})=0$), or no measurable spatial extent ($\upmu_3(\delta \Updelta^{(n)}_t)=0$), or both.
In practical terms, a hap whose manifestation is so small that falls below the instruments' uncertainty or simulation error, will be considered negligible and thus non-occurring.
In the following, we shall use the expression ``Except, Perhaps, in a Finite Number of Realisations'' (EPiaFNoR), 
to denote that something occurs in almost all realisations, but the occasional non-occurrence, 
being only a finite number of times, has no influence in any mean field or quantity.
On the other hand, when a hap occurs in a denumerable infinite number of realisations, it might or not be transmitted to the mean fields,
depending upon the relative size of such an infinite number.
The following definition will help to assess the transfer from realisations to mean fields:
\begin{definition}
\label{def:lesserInf}
Let $n\in \mathbb{N}$ be the index of the ensemble $\mathfrak{E}$.
Let $f(n)$ and $g(n)$ be two functions of the index $n$ such that 
\[ \lim \limits_{n \rightarrow \infty} f(n) = \infty  \ ; \quad \lim \limits_{n \rightarrow \infty} g(n) = \infty    \]
A function $f(n)$ is a \textbf{lesser infinite} than $g(n)$ iff
\[ \lim \limits_{n \rightarrow \infty} \frac{f(n)}{g(n)} =0  \] 
\end{definition}

Typically, this definition will be used with the quotient $f(n)/n$, as observed below.
For example, if a hap occurs with a frequency $f(n)=\sqrt{n}$ in the ensemble, then it would be a lesser infinite than the number of realisations.
Therefore, it is pertinent to characterise how frequently does a hap occur in the ensemble $\mathfrak{E}$, which is done below.
\begin{definition}
 \label{def:hapFreq}
 Let $\mathfrak{E}$ be an ensemble of $N$ realisations identified by the index $n \in \mathbb{N}$, $n=1,2, \ ...\ N$.
 Assume that a hap $\mathfrak{H}$ occurs $K \leq N$ times in the ensemble, each occurrence identified by an index $k \in \mathbb{N}$, $k=1,2, \ ...\ K$.
The hap may occur at different times and positions, but it is positively identified as the same hap $\mathfrak{H}$ in all $K$ occasions.
Then, the \textbf{frequency} of the hap $\mathfrak{H}$ in the ensemble $\mathfrak{E}$ is defined as:
\begin{equation}
 \mathfrak{H}_f=\lim \limits_{N \rightarrow \infty} \frac{K}{N}
\end{equation}
The hap $\mathfrak{H}$ occurs in every realisation EPiaFNoR iff $\mathfrak{H}_f=1$.
Haps with $\mathfrak{H}_f=1$ are called \textbf{persistent}.
\end{definition}

If $0 \leq \mathfrak{H}_f<1$ then the hap is not occurring in every realisation. 
For example, $\mathfrak{H}_f=\tfrac15$ means that, on average, the hap occurs once every five realisations EPiaFNoR.
Furthermore, if the hap occurs in a lesser infinite of realisations, that is, if $K$ is a lesser infinite than $N$, then $\mathfrak{H}_f=0$ despite the hap still occurring an infinite number of times in the ensemble.
It follows that infrequent haps have it difficult to be transmitted into the ensemble-averaged fields, 
because the division by $N$ in Eqs. \eqref{eq:ensembleAve}-\eqref{eq:ensembleAveChiralR} would iron out any contribution that is not repeatable enough.
This line of thinking leads to the following definition:
\begin{definition}
A hap $\mathfrak{H}$ is \textbf{sparse} in the ensemble $\mathfrak{E}$ iff $\mathfrak{H}_f=0$.
\end{definition}
It also leads to the conclusion that a sparse hap $\mathfrak{H}$ is not transmitted to the ensemble-averaged fields, as stated in the following proposition:
\begin{theorem}
\label{thr:sparse}
 Let $\mathfrak{H}$ be a sparse hap in the ensemble $\mathfrak{E}$ and $\Psi(t,\textbf{x})$ the corresponding ensemble-averaged field obtained from Eqs. \eqref{eq:ensembleAve}-\eqref{eq:ensembleAveChiralR}.
 Let $\Psi^-(t,\textbf{x})$ be the ensemble-averaged field that would be obtained from Eqs. \eqref{eq:ensembleAve}-\eqref{eq:ensembleAveChiralR} if the hap $\mathfrak{H}$ were non-existent in any realisation, called the hapless mean field.
Then, $\Psi(t,\textbf{x})=\Psi^-(t,\textbf{x})$, $\forall \ (t,\textbf{x}) \in \Upomega$.
\end{theorem}
\begin{proof}
 Assume an ensemble $\mathfrak{E}$ of $N$ realisations, identified by the index $n=1,2,...,N$, in which the hap $\mathfrak{H}$ occurs $K$ times, identified with the index $k=1,2,...,K$,
 and it does not occur $N-K$ times, identified with the index $m=1,2,...,N-K$. 
 Let $\psi^{(m)}(t,\textbf{x})$ be the physical field in a realisation in which the hap does not occur and  $\psi^{(k)}(t,\textbf{x})$ the physical field in a realisation in which it occurs.
 Assume $\psi^{(k)}$ can be written as $\psi^{(k)}(t,\textbf{x})=\psi^{(k)-}(t,\textbf{x})+\psi_h^{(k)}(t,\textbf{x})$, Eq. \eqref{eq:hapDecomp}, 
 where $\psi^{(k)-}(t,\textbf{x})$ is the hapless field, while $\psi_h^{(k)}(t,\textbf{x})$ is the hap field.
 Note that: (i) in a realisation in which the hap does not occur $\psi^{(m)}(t,\textbf{x})\equiv \psi^{(m)-}(t,\textbf{x})$ and $\psi_h^{(m)}(t,\textbf{x})=0$; 
 (ii) in a realisation in which the hap occurs, the hapless field $\psi^{(k)-}(t,\textbf{x})$ would not be truly a physical field, albeit it would be analogous to a true physical field like $\psi^{(m)}(t,\textbf{x})$;
 and (iii) the hapless field $\psi^{(n)-}(t,\textbf{x})$ exists in every realisation, either as in (i) or as in (ii).
 The hapless mean field is then defined as (or equivalent equations to Eqs. \eqref{eq:ensembleAve}-\eqref{eq:ensembleAveChiralR}):
 \begin{equation}
  \label{eq:haplessMean}
  \Psi^-(t,\textbf{x})=\lim \limits_{N\rightarrow \infty} \frac{1}{N} \sum \limits_{n=1}^{N} \psi^{(n)-}(t,\textbf{x})
 \end{equation}
 Let $M^{(k)}>0$ be the maximum of $|\psi_h^{(k)}(t,\textbf{x})|$, i.e., $|\psi_h^{(k)}(t,\textbf{x})| \leq M^{(k)}$, $\forall \ (t,\textbf{x}) \in \Upomega$.
  $M^{(k)}$ exists and is finite because $\psi_h^{(k)}(t,\textbf{x}) \in H^2(\Upomega)$ is bounded.
 It follows that $\psi^{(k)}(t,\textbf{x}) \leq \psi^{(k)-}(t,\textbf{x})+ M^{(k)}$, $\forall \ (t,\textbf{x}) \in \Upomega$.
 Let $M^{(+)}=\sup \left\{ M^{(k)}  \right\}<+\infty$ be the maximum of all $ M^{(k)}$.
 Substituting $\psi^{(k)}$ in Eq. \eqref{eq:ensembleAve} yields:  
 \begin{align*}
  & \frac{1}{N} \sum \limits_{n=1}^{N}  \psi^{(n)}(t,\textbf{x}) = \frac{1}{N} \left( \sum \limits_{m=1}^{N-K}  \psi^{(m)} + \sum \limits_{k=1}^K  \psi^{(k)}  \right )=
 \frac{1}{N} \left( \sum \limits_{m=1}^{N-K}  \psi^{(m)} + \sum \limits_{k=1}^K \left( \psi^{(k)-}+\psi_h^{(k)} \right)  \right )=  \\
 &\stackrel{(1)}{=} \frac{1}{N} \left( \sum \limits_{n=1}^{N}  \psi^{(n)-} + \sum \limits_{k=1}^K \psi_h^{(k)}  \right ) \leq \frac{1}{N} \sum \limits_{n=1}^{N}  \psi^{(n)-} +  \frac{1}{N}  \sum \limits_{k=1}^K M^{(k)}   
   \leq  \frac{1}{N} \sum \limits_{n=1}^{N}  \psi^{(n)-} +  \frac{K}{N}  M^{(+)}  \Rightarrow \\
 &\frac{1}{N} \sum \limits_{n=1}^{N} \left(  \psi^{(n)} - \psi^{(n)-}  \right) \stackrel{(2)}{=} \frac{1}{N} \sum \limits_{k=1}^K \left(  \psi^{(k)} - \psi^{(k)-} \right) \leq \frac{1}{N} \sum \limits_{k=1}^K |  \psi^{(k)} - \psi^{(k)-}| = \frac{1}{N} \sum \limits_{k=1}^K |  \psi_h^{(k)}|   \leq  \frac{K}{N}  M^{(+)}  \Rightarrow \\
 &\lim \limits_{N\rightarrow \infty} \frac{1}{N} \sum \limits_{k=1}^K | \psi_h^{(k)}| \leq \lim \limits_{N\rightarrow \infty}  \frac{K}{N}  M^{(+)} \stackrel{(3)}{=} 0 \Rightarrow \lim \limits_{N\rightarrow \infty} \frac{1}{N} \sum \limits_{k=1}^K | \psi_h^{(k)}(t,\textbf{x}) |=0 
 \end{align*}
The steps with numerals are explained next.

 (1): The sum in $n$ leads to the definition of $\Psi^-(t,\textbf{x})$ and includes both types of fields: The physical fields $\psi^{(m)}(t,\textbf{x})\equiv \psi^{(m)-}(t,\textbf{x})$, truly without hap, and the hapless fields $\psi^{(k)-}(t,\textbf{x})$, 
 resulting from removing the hap contribution. 
 It is written as $\psi^{(n)-} $ to distinguish it from the first sum in $n$ with $\psi^{(n)}$, although the fields $\psi^{(m)} $ are identical in both sums in $n$.
 
 (2): The physical fields $\psi^{(m)}$ cancel identically in the sum in $n$, remaining only the terms in $k$, in which $\psi^{(k)}$ is a true physical field and $\psi^{(k)-}$ is the hapless field.
 Their difference is the hap field $\psi_h^{(k)}$.
 
 (3): Since the hap is sparse, by definition it must be $K/N\rightarrow 0$ as $N\rightarrow \infty$.
 Recall $ M^{(+)}$ is finite.
 
 The initial sum in $n$ with $\psi^{(n)}$ leads to $\Psi(t,\textbf{x})$ by definition. 
Likewise, the sum in $n$ with $\psi^{(n)-}$ leads to $\Psi^-(t,\textbf{x})$.
The last equality shows that the difference between $\Psi$ and $\Psi^-$ vanishes as $N\rightarrow \infty$, because it is a lesser infinite than $N$, which proves the theorem.
 \end{proof}

Sparse haps are not transmitted to the mean fields, regardless of how impressive they may seem in those realisations in which they occur.
Thus, they are not particularly interesting for the present research.
But nothing has been said yet about non-sparse haps.
We shall next develop a method to quantify how much is a hap being transmitted to the mean fields.
The method has into account that haps have finite size and duration, and that they not always occur at the same place nor at the same time.
These notions are defined now.
\begin{definition}
 \label{def:hapSize}
 Let $\mathfrak{E}$ be an ensemble of $N$ realisations, characterised by the index $n=1,2,...,N$.
 Assume that a hap $\mathfrak{H}$ occurs $K \leq N$ times in the ensemble and the spacetime domain of the hap in the $n^{th}$ realisation is $\mathfrak{H}_{\upomega}^{(n)} \subset \Upomega$,
 $\upmu(\mathfrak{H}_{\upomega}^{(n)}) \geq 0$, with $\upmu:\Upomega \rightarrow  \{0 \} \cup \mathbb{R}^+$ the canonical measure in $\mathbb{R}^4$.
 Let $k=1,2,...,K$ be the index identifying the realisations in which the hap occurs, i.e.,  $\upmu(\mathfrak{H}_{\upomega}^{(k)})>0$, $\forall \ k=1,2,...,K$.
The \textbf{intrinsic mean size}
\footnote{In this research, an \textit{intrinsic mean} is an averaging operation in which all $K$ terms are significant, i.e., zero or negligible terms are excluded from the sum, which is then divided by the number of significant terms $K$.
In this particular example, haps with $\upmu(\mathfrak{H}_{\upomega}^{(k)}) \approx 0$, that is, with  $\upmu(\mathfrak{H}_{\upomega}^{(k)})< \epsilon'$ for given small $\epsilon'>0$, are considered not significant.
On the contrary, a \textit{mean} is an averaging operation in which not all $N$ terms of the sum need be significant, which is then divided by $N$.
Intrinsic means are denoted with a hat $\hat{\cdot}$ .} 
of the hap $\mathfrak{H}$ in the ensemble $\mathfrak{E}$ is defined as:
\begin{equation}
\label{eq:intrinsicMeanB}
 \hat{\upmu}(\mathfrak{H}):= \frac{1}{K} \sum \limits_{k=1}^{K}  \upmu(\mathfrak{H}_{\upomega}^{(k)}) 
\end{equation}
if $K$ is bounded ($K>0$ and $\lim \limits_{N \rightarrow \infty} K < \infty$), or 
\begin{equation}
\label{eq:intrinsicMeanU}
 \hat{\upmu}(\mathfrak{H}):=\lim \limits_{K \rightarrow \infty} \frac{1}{K} \sum \limits_{k=1}^{K}  \upmu(\mathfrak{H}_{\upomega}^{(k)})
\end{equation}
if $K$ is unbounded ($\lim \limits_{N \rightarrow \infty} K = \infty$).
If $K=0$ then $ \hat{\upmu}(\mathfrak{H})\equiv 0$.

Alternatively, the \textbf{mean size} of the hap is defined as:
\begin{equation}
\label{eq:meanSize}
 \upmu(\mathfrak{H}) :=\lim \limits_{N \rightarrow \infty} \frac{1}{N} \sum \limits_{n=1}^{N}  \upmu(\mathfrak{H}_{\upomega}^{(n)}) 
\end{equation}
 The series in Eqs. \eqref{eq:intrinsicMeanU} \& \eqref{eq:meanSize} converge absolutely because $\Upomega$ is a bounded domain ($\upmu(\Upomega)<\infty$).
 It is verified that ${\upmu}(\mathfrak{H}) \leq \hat{\upmu}(\mathfrak{H}) \leq \upmu(\Upomega)$.
\end{definition}
Note that neither $ \hat{\upmu}(\mathfrak{H})$ nor ${\upmu}(\mathfrak{H})$ contain information about the spacetime localisation of the hap, only about its mean size.
The following proposition is obvious and is offered without demonstration:
\begin{lemma}
 Let $\mathfrak{H}$ be a sparse hap.
 Then ${\upmu}(\mathfrak{H})=0$ and $\hat{\upmu}(\mathfrak{H})>0$.
\end{lemma}
If $\hat{\upmu}(\mathfrak{H})\mathop{=}0$, then it is understood that the hap is non-existent in the ensemble, because $K\mathop{=}0$.
If the hap is significant in at least one realisation, then $K>0$, $\upmu(\mathfrak{H}_{\upomega}^{(k)})>0 $ for some $k$, and $\hat{\upmu}(\mathfrak{H})>0$.
It is obvious that if $K$ is bounded, then the hap is necessarily sparse.
Henceforth, only haps with unbounded $K$ will be considered.

The hap $\mathfrak{H}$ may occur at different times and spatial locations in different realisations, i.e., 
the hap's centroid would change noticeably from one realisation to the other, $\textbf{x}_{\mathfrak{H}}^{(n)}(t) \not \approx \textbf{x}_{\mathfrak{H}}^{(m)}(t)$, $n \neq m$.
How would this affect the mean fields $\Psi(t,\textbf{x})$, in which each averaging operation must be executed at fixed $(t,\textbf{x}) \in \Upomega$?
Since $\Upomega$ is bounded, the hap's size is also bounded.
Actually, for any hap the ratio $0<\hat{\upmu}(\mathfrak{H})/\upmu(\Upomega)\leq 1$ holds, and likewise for any non-sparse hap it is $0<\upmu(\mathfrak{H})/\upmu(\Upomega)\leq 1$.
It follows that, for $N\rightarrow \infty$, a non-sparse hap must in the end repeat indefinitely its occurrence at every available place and time interval,
just as a non-skilled shooter will at the end hit several times every point in the target, including the bullseye, after a million shots.
In other words, there exist infinite realisations pairs $n$ and $m$ such that 
$\mathfrak{H}_{\upomega}^{(n)} \cap \mathfrak{H}_{\upomega}^{(m)} \neq \emptyset$.
Let $q=\lceil \upmu(\Upomega) / \upmu(\mathfrak{H}) \rceil \in \mathbb{N}$ ($\lceil \cdot \rceil$ is the ceiling function), where typically $q\gg1$. 
At the very latest, every $q$ realisations must the haps have a non-empty intersection, with non-zero measure, $\upmu(\cap_1^q \mathfrak{H}_{\upomega}^{(n)})\neq 0$,
which means that there must exist a spacetime domain in which the hap repeats indefinitely as $N\rightarrow \infty$,
$\cap_n  \mathfrak{H}_{\upomega}^{(n)}\neq \emptyset $.
Therefore, a non-sparse hap is always affecting the ensemble-averaged fields of the flow, although it may affect them very little (in a worst-case scenario, the influence would be that of a typical realisation divided by $q$).
If $N$ remains finite, it may well occur that $\cap_n  \mathfrak{H}_{\upomega}^{(n)} \mathop{=} \emptyset $, because the hap always occurs at different locations or intervals,
in which case the influence on the mean fields would be negligible.
We purport to quantify the effect of haps in the mean fields.

\begin{definition}
 Let $\mathfrak{H}$ be a hap in the ensemble $\mathfrak{E}$.
 The \textbf{total spread} of the hap is defined as:
 \begin{equation}
 \label{eq:totalSpread}
  \mathfrak{H}_s:=\lim \limits_{N \rightarrow \infty} \frac{\upmu \left( \bigcup \limits_{n=1}^N  \mathfrak{H}_{\upomega}^{(n)} \right) }{\upmu(\Upomega) }
 \end{equation}
  with ${\upmu(\mathfrak{H}})/ {\upmu(\Upomega)} \leq \mathfrak{H}_s \leq 1$.
  
 The hap is \textbf{nomadic} or \textbf{fully scattered} if $\mathfrak{H}_s=1 $, i.e., the hap ends up occupying all available space.
 The hap is \textbf{fixed} or \textbf{fully localised} if 
 \begin{equation}
 \label{eq:fullyLocalised}
  \mathfrak{H}_s=\frac{ \upmu(\mathfrak{H})}{ \upmu(\Upomega)}
 \end{equation}
\end{definition}

Note that even uncommon realisations with oddly placed haps contribute to the numerator in Eq. \eqref{eq:totalSpread}, hence its name of `\textit{total spread}'. 
However, as long as $N$ remains finite, the probability of such an ill-behaved hap might be negligible.
From Eqs. \eqref{eq:meanSize}, \eqref{eq:totalSpread} \& \eqref{eq:fullyLocalised} it follows for fixed haps:
\[ \upmu(\mathfrak{H}) := \lim \limits_{N \rightarrow \infty} \frac{1}{N} \sum \limits_{n=1}^{N}  \upmu(\mathfrak{H}_{\upomega}^{(n)}) = \lim \limits_{N \rightarrow \infty} \upmu \left( \bigcup \limits_{n=1}^N  \mathfrak{H}_{\upomega}^{(n)} \right)  \]
which for finite $N$ implies two approximate results: \(  \upmu(\mathfrak{H}_{\upomega}^{(n)}) \approx \upmu(\mathfrak{H}) \) and 
\(  \upmu(\mathfrak{H}_{\upomega}^{(m)}) \approx \upmu  \left( \bigcup \limits_{n=1}^N  \mathfrak{H}_{\upomega}^{(n)} \right), \ 1 \leq m \leq N  \).
The first result means that all haps have almost the same size, which is equal to the hap's mean size.
The second means that all haps occur approximately in the same spacetime domain, since the union of all $N$ haps is not substantially greater than any of the union's components.
Otherwise put, the centroids $\textbf{x}_{\mathfrak{H}}^{(n)}(t)$ of different realisations are very near one another, $\textbf{x}_{\mathfrak{H}}^{(n)}(t) \approx \textbf{x}_{\mathfrak{H}}^{(m)}(t)$, $n \neq m$.  
The total spread of a hap is an indicator of its degree of transmission to the mean fields.
A localised hap is more intensely transmitted than a scattered hap.

Another global parameter influencing the hap's degree of transmission to the ensemble average, closely related to $\mathfrak{H}_s$, is the following:
\begin{definition}
 Let $\mathfrak{H}$ be a hap in the ensemble $\mathfrak{E}$.
 The \textbf{mean confinement} of the hap is defined as:
 \begin{equation}
 \label{eq:meanConfine}
  \mathfrak{H}_c :=\lim \limits_{N \rightarrow \infty} \frac{\upmu(\mathfrak{H})} {\upmu \left( \bigcup \limits_{n=1}^N  \mathfrak{H}_{\upomega}^{(n)} \right) }, \quad 
 \end{equation}
 with ${\upmu(\mathfrak{H})}/ {\upmu(\Upomega)} \leq \mathfrak{H}_c \leq 1$.
 
 The hap is {fixed} if $\mathfrak{H}_c=1 $, while the hap is {nomadic} if 
 \begin{equation}
 \label{eq:fullyScattered}
  \mathfrak{H}_c=\frac{ \upmu(\mathfrak{H})}{ \upmu(\Upomega)}
 \end{equation}
\end{definition}
Again, even uncommon realisations of the hap with unusual spacetime domains $\mathfrak{H}_{\upomega}^{(n)}$ will contribute to Eq. \eqref{eq:meanConfine}.

Finally, a parameter that characterises the global degree of transmission of a hap to the ensemble-averaged fields:
\begin{definition}
 Let $\mathfrak{H}$ be a hap in the ensemble $\mathfrak{E}$.
 The \textbf{global repeatability} of the hap $\mathfrak{H}$ is defined as the product of its frequency by its mean confinement:
\begin{equation}
 \label{eq:globalRepeat}
 \mathfrak{H}_r =   \mathfrak{H}_f  \  \mathfrak{H}_c
\end{equation}
\end{definition}
A persistent ($\mathfrak{H}_f= 1$) and fixed ($\mathfrak{H}_c = 1$) hap occurs in every realisation EPiaFNoR, and always in the same position and time interval.
It is thus a fully repeatable hap that gets maximally transmitted to the mean fields ($ \mathfrak{H}_r =1$). 
An infrequent ($\mathfrak{H}_f \ll 1$) and scattered ($\mathfrak{H}_c \ll 1$) hap would be poorly repeatable ($\mathfrak{H}_r \ll 1$), 
and its degree of transmission to the mean fields would be quite low.
It might well occur that such a sparingly repeatable hap falls below the uncertainty of measuring instruments or simulation resolution.
In such cases, from a practical point of view, the hap would not be transmitted to the ensemble average, although it may have significant values in the few realisations in which it occurs. 

The above quantities characterising a hap are global, i.e., a single number quantifies how much is the hap transmitted to the mean fields.
However, it is possible a much more accurate quantification, which takes the form of a mean scalar field that determines the hap's repeatability at each spacetime point.
This scalar field uses the indicator function of the hap's domain $\mathfrak{H}_{\upomega}^{(n)}$ in every realisation of the ensemble, $\mathbb{1}_{\mathfrak{H}_{\upomega}^{(n)}}(t,\textbf{x})$, Eq. \eqref{eq:indicatorFunct},
regardless of the hap's occurrence.
\begin{definition}
	Let $\mathfrak{H}$ be a hap in the ensemble of realisations $\mathfrak{E}$.
	The hap's \textbf{index of repeatability} is defined as the mean scalar field:
	\begin{equation}
		\uprho_{\mathfrak{H}}(t,\textbf{x}) := \lim \limits_{N\rightarrow \infty} \frac{1}{N} \sum \limits_{n=1}^N \mathbb{1}_{\mathfrak{H}_{\upomega}^{(n)}}(t,\textbf{x}), \quad (t,\textbf{x})\in \Upomega
	\end{equation}
	The hap is \textbf{fully repeatable} in a neighbourhood of $(t,\textbf{x})$ if $\uprho_{\mathfrak{H}}(t,\textbf{x})\mathop{=}1$, 
	and is \textbf{fully non-repeatable} in a neighbourhood of $(t,\textbf{x})$ if $\uprho_{\mathfrak{H}}(t,\textbf{x})\mathop{=}0$.
	A hap occurring at $(t,\textbf{x})$ a number of times $K$ that is a lesser infinite than $N$ satisfies $\uprho_{\mathfrak{H}}(t,\textbf{x})\mathop{=}0$. 
\end{definition}
Recalling now the notions of hapless field, $\psi^{(n)-}(t,\textbf{x})$, and hap field, $\psi_h^{(n)}(t,\textbf{x})$, introduced in Eq. \eqref{eq:hapDecomp}, 
it is possible to express the mean field of a flow, $\Psi(t,\textbf{x})$,
with the help of the index of repeatability, $\uprho_{\mathfrak{H}}(t,\textbf{x})$.
Such an expression takes the form:
\begin{equation}
 \label{eq:GarciaDecomp}
 \Psi(t,\textbf{x})=\Psi^-(t,\textbf{x}) + \uprho_{\mathfrak{H}}(t,\textbf{x}) \hat{\Psi}_h(t,\textbf{x})
\end{equation} 
In Eq. \eqref{eq:GarciaDecomp}, $\Psi(t,\textbf{x})$ is defined generally by Eqs. \eqref{eq:ensembleAve}-\eqref{eq:ensembleAveChiralR}, $\Psi^-(t,\textbf{x})$ by Eq. \eqref{eq:haplessMean} and 
$ \hat{\Psi}_h(t,\textbf{x})$ is the intrinsic-mean hap field, given by:
\begin{equation}
 \label{eq:hapMean}
 \hat{\Psi}_h(t,\textbf{x}) = \lim \limits_{N \rightarrow \infty} \frac{1}{K} \sum \limits_{k=1}^K \psi_h^{(k)}(t,\textbf{x})
\end{equation}
whereby each term $\psi_h^{(k)}(t,\textbf{x})$ in the sum is significant, i.e. the index $k=1,2,...,K$ only counts instances of non-zero (or non-negligible) hap fields at $(t,\textbf{x})$. 
The researcher that observes haps in some realisations, only needs to isolate the hap fields (whenever and wherever they occur), 
calculate the intrinsic mean and apply Eq. \eqref{eq:GarciaDecomp} to determine $\Psi(t,\textbf{x})$ approximately
(we say \textit{approximately} because, rigorously, the researcher should perform infinite realisations).

In summary, whenever a researcher performs an experiment or a computer simulation (i.e., a realisation), whatever interesting feature that occurs in it should be assessed for repeatability in the following experiments or simulations.
Only features repeatable enough will be transmitted to the mean fields, {repeatable} meaning a combination of frequency and spread.
The study of haps just sketched furnishes a line of work that other researchers may explore or enhance.
It should be clear now that the relationship between individual realisations and ensemble-averaged fields is far from being straightforward.
The study of haps initiated herein may be completed with notions such as the hap's barycentre in each realisation, the barycentric radius (which is the radius of the minimum 4-sphere than contains all barycentres of the ensemble),
the barycentric streamline in each realisation, etc., which are beyond the scope of the present research.

\end{document}